%% file: paper.tex
\documentclass[sigconf]{./sig-alternate-10pt}

\usepackage[
pass,
]{geometry}

\usepackage[ruled]{algorithm2e}
\usepackage{graphicx}
\usepackage{amsmath, amssymb}
\usepackage{booktabs}
\usepackage{subfigure}
\usepackage{algorithmic}
\usepackage{paralist}
\usepackage{multirow}
\usepackage{enumitem}
\usepackage[english]{babel}
\usepackage{hyperref}
\usepackage{color, colortbl}
\usepackage{cite}
\usepackage{tabularx}
\usepackage{multirow}
\usepackage{balance}

\newcommand{\controller}{MSP}
\newcommand{\js}{JavaScript}
\newcommand{\webmining}{web-mining}
\newcommand{\testbed}{WebTestbench}
\definecolor{Gray}{gray}{0.9}
\definecolor{LightCyan}{rgb}{0.88,1,1}

\hypersetup{
  bookmarks=true, 
  colorlinks=true,%
  citecolor=black,%
  filecolor=black,%
  linkcolor=black,%
  urlcolor=black,%
  pagebackref=true,%
}

\definecolor{mygreen}{rgb}{0,0.6,0}
\definecolor{mygray}{rgb}{0.5,0.5,0.5}
\definecolor{mymauve}{rgb}{0.58,0,0.82}

\def\ttfntsize{9}
\makeatletter
\let\oldtexttt\texttt
\let\texttt\@undefined
\makeatother
\newcommand{\texttt}[1]{\fontsize{\ttfntsize}{\ttfntsize}\oldtexttt{#1}}
\makeatletter
\let\oldtt\tt
\let\tt\@undefined
\makeatother
\newcommand{\tt}{\fontsize{\ttfntsize}{\ttfntsize}\oldtt} 

\definecolor{Gray}{gray}{0.9}
\definecolor{LightCyan}{rgb}{0.88,1,1}

\begin{document}
\title{Truth in Web Mining: Measuring the Profitability and Cost of Cryptominers as a Web Monetization Model}

\numberofauthors{3}
\author{
\alignauthor
Panagiotis Papadopoulos\\
\affaddr{FORTH-ICS, Greece}\\
\email{panpap@ics.forth.gr}
\and
\alignauthor
Panagiotis Ilia\\
\affaddr{FORTH-ICS, Greece}\\
\email{pilia@ics.forth.gr}
\alignauthor
Evangelos P. Markatos\\
\affaddr{FORTH-ICS, Greece}\\
\email{markatos@ics.forth.gr}
}


\maketitle

\input{abstract}
\keywords{Cryptomining, Cost of Web-mining, Digital Advertising, Cryptocurrency, Cryptojacking}
\input{introduction}
\input{background}
\input{dataset}
\input{measurements}
\input{discussion}

\input{related}
\input{summary}

\balance
\bibliographystyle{abbrv}
\bibliography{paper}

\end{document}

%% file: abstract.tex
\begin{abstract}
The recent advances of web-based cryptomining libraries along with the whopping market 
value of cryptocoins have convinced an increasing number of publishers to switch to web mining
as a source of monetization for their websites. The conditions
could not be better nowadays: the inevitable arms race between adblockers and 
advertisers is at its peak with publishers caught in the crossfire.
But, \emph{can cryptomining be the next primary monetization model in the post advertising
era of free Internet?}

In this paper, we respond to this exact question. In particular, we compare the 
profitability of cryptomining and advertising to assess the most advantageous option 
for a content provider. In addition, we measure the costs imposed to the user in each case 
with regards to power consumption, resources utilization, network traffic, device temperature and user experience.
Our results show that cryptomining can surpass the profitability of advertising under 
specific circumstances, however users need to sustain a significant cost on their devices.
\end{abstract}

%% file: introduction.tex
\section{Introduction}
\label{sec:introduction}

Digital advertising is today the dominant monetization model for web publishers. 
During the last decade, it has become the driving force of 
the web, leading to the provision and support of new web services and 
applications~\cite{followTheMoney,breaking}. 
Indicatively, digital advertising, which is continuously growing with an unprecedented rate, reached total revenues of \$209 billion 
in 2017~\cite{digital_ad_vs_tv}.

However, in the recent years, either due to the roaring privacy implications of targeted advertising~\cite{razaghpanah2018apps,vallina2016tracking,carrascosa2015always} or the irritation dodgy ads may cause~\cite{Goldstein:2013:CAA:2488388.2488429}, a growing number of users (615 million devices -- 30\% growth since last year~\cite{adblockingUsers}) decided to abdicate from receiving ads by adopting all-out approaches (like deploying ad-blocking mechanisms~\cite{qhostery,adblockplus,appsVsWeb} or ad-stripping browsers~\cite{cliqz,brave,ChromeBlocker}).
This increasing ad-blocking trend made some major web publishers, after seeing their income to significantly shrink 
(total losses of \$22 billion~\cite{revenueDrop}), to deploy ad-blocker detection techniques~\cite{mughees2017detecting,nithyanand2016adblocking,iqbal2017ad} and deny serving content to 
ad-blocking users~\cite{wired,forbes,financialTimes,antiadblock}.
Such aggressive actions from both sides escalated an inevitable arms race between the ad-ecosystem on the one side, 
and the ad-blockers and privacy advocates on the other side~\cite{armsRace,nithyanand2016adblocking,extortion}.

It is of no doubt that in such a dispute, publishers were trapped in the crossfire being unable to effectively monetize their services.
To that end, it did not take long for some of them to look for effective and reliable alternative schemes to support their websites.
Some of these schemes include paid website versions, user compensation (e.g., Basic Attention Token~\cite{attensionToken}) and cryptomining.
Especially the latter, given its privacy protecting nature (no user tracking and personal data collection required, thus 
making cryptocurrency mining GDPR compliant) and the frenetic increase of the market value of cryprocoins, gains an ever increasing popularity.

Of course, in-browser based mining is not a new idea. The compatibility of Javascript miners with all modern browsers gave motivation 
for web mining attempts since the very early days of Bitcoin, back in 2011~\cite{bitcoinPlus}.
To that end, web miners ``borrow'' spare CPU cycles of the visiting users' devices for 
performing their Proof-of-Work (PoW) computations~\cite{pow} for as long as the user is browsing the website's content. 
However, the increased mining difficulty of Bitcoin was the primary factor that led such approaches to failure.
Yet, the rapid growth of Bitcoin lured several initiatives to construct their own derivatives 
(more than 1638 nowadays~\cite{altcoins}) providing specific extra features e.g., transaction speed, proof-of-stake. 
One of the provided features: mining speed, became the growth factor for some coins (Monero~\cite{monero,moneroHashRate} 
grew from 13\$ to 300\$ within 2017~\cite{moneroGrowth}) and worked as catalyst
a for the incarnation of web cryptomining~\cite{miningSites}.

Indeed, since the release of the first \js\ miner (i.e., September 2017) by Coinhive~\cite{coinhive-api}, 
we observe a rapidly increasing~\cite{mining_trend,mineFever,cryptojackingPaper} number of content providers deploying web-based 
cryptomining libraries in their websites, monetizing their content either by using both ads and \webmining\ or by fully replacing ads 
(e.g. PirateBay~\cite{piratebay}). 
So the important question that arises at this point is the following: \emph{Can \webmining\ become the next business model 
of the post ad-supported era of Internet?}

There are numerous opinions about this subject~\cite{debate,webminingwaste,incomeExample}, but it is apparent that in order to accurately respond to such a question we first 
need to investigate all aspects of both advertising and \webmining. These aspects include, first of all, the profitability that cryptominers provide to publishers and
also the costs that users have to sustain from the utilization of their resources: let us not forget that the unsustainable costs~\cite{www18adcost,truthInAdvertising} 
of advertising made ad-blocking popular.

In this study, we aim to address exactly that; we conduct the first full-scale analysis of the profitability and costs of \webmining, 
in an attempt to shed light in the newly emerged technology of in-browser cryptomining and explore if it can replace ads on the web.
Specifically, in this study we estimate the possible revenues for the different monetization strategies: advertising and \webmining, 
aiming to determine under what circumstances a miner-supported website can surpass the profits from digital advertising.

Additionally, we collect a large dataset of miner- and ad- supported websites 
and by designing and developing \testbed, a sensor-based testbed, we measure 
the resource
utilization of both models in an attempt to compare their imposed user-side costs. In particular, \testbed\ is capable of measuring (i) the utilization 
of mining regarding system resources such as CPU and main memory, (ii) the degradation of the user experience
due to the increased mining workload, (iii) the energy consumption and how this affects battery-operated devices (e.g., laptops, tablets, smartphones), 
(iv) system temperature and how overheating affects the user's device and (v) network and how this can affect a possible mobile dataplan.

To summarize, in this paper we make the following contributions:

\begin{enumerate}
	\item We conduct the first study on the profitability of web-based cryptocurrency mining, questioning the ability of mining to become a reliable monetization method for future web services. Our results show that for the average duration of a website visit, ads are
	5.5x more profitable than cryptomining. However, a miner-supported website can produce higher revenues if the visitor remains in the website for longer than 5.3
	minutes.
	
	\item We design a methodology to assess the resource utilization patterns of ad- and miner- supported websites on the visitor's device. We implement our 
	approach in \testbed\ framework and we investigate what costs these utilization patterns impose on the visitor's side with regards to the user experience, the
	system's temperature, and energy consumption and battery autonomy.
	
	\item We collect a large dataset of around 200K ad- and miner-supported websites that include different \webmining\ libraries and cryprocurrencies. We use this dataset as input for the 
	\testbed\ framework and we compare the resource utilization and costs of the two web monetization models. Our results show that while browsing a miner-supported website, the visitor's CPU gets utilized 59 times more 
	than while visiting an ad-supported website, thus increasing the 
	temperature (52.8\%) and power consumption (2x) of her device.
	
\end{enumerate}

%% file: background.tex
\section{Background}
\label{sec:background}

\subsection{Web-based cryptocurrency mining}
Web-based mining is a method of cryptocurrency mining that 
happens inside a browser, using a mining script delivered 
through a website.
The idea of in-browser cryptocurrency mining gained 
popularity since the very early days of cryptocurrency (i.e., Bitcoin), 
back in 2011. Indeed, the popularity of web applications and the 
wide compatibility of \js\ based miners with all modern browsers and 
platforms drew the attention of the developers of 
BitcoinPlus~\cite{bitcoinPlus}. However, the increased mining 
difficulty of Bitcoin not only made web-based mining 
unprofitable but, practically, infeasible. 

Few years later, the rise of alternative cryptocoins (altcoins) that provide distributed mining, increased mining speed and ASIC (Application-Specific Integrated Circuit)
resistance, made distributed CPU (i.e, x86, x86-64, ARM) based mining effective~\cite{cpuonly,monero}, even when using commodity hardware. 
As a consequence, all these new altcoins, such as Electroneum, Sumokoin, 
Bytecoin and Monero, not only revived the concept of in-browser cryptomining 
but also opened new funding avenues for web publishers. 

The motivation behind this new business model is simple: 
users visit a website and pay for the received content indirectly 
by mining cryptocurrency coins, without being polluted with (possibly annoying~\cite{Goldstein:2013:CAA:2488388.2488429}) ads. Furthermore, publishers do not have to bother collecting behavioral data, including trackers~\cite{pujol} or 
user fingerprinting libraries~\cite{Nikiforakis:2013:CME:2497621.2498133} to get higher prices~\cite{imcRTB} for their ad-slots. 
As a consequence, users get a cleaner, faster, and potentially less risky~\cite{malvertising,Zarras:2014:DAM:2663716.2663719} website. 


\subsection{Monero crypto-coin}
Monero, is the most popular altcoin for \webmining\ at the moment, 
growing from 13\$ to 300\$ within 2017~\cite{moneroGrowth}. 
It is based on the CryptoNight Proof-of-Work (PoW) hash algorithm~\cite{cryptonight}, which comes from the CryptoNote protocol. The motivation behind the development of Monero was to provide decentralization and privacy by obscuring the sender, recipient and amount 
of every transaction made.
Although Monero is ASIC resistant, 
and thus can be mined with both CPUs and GPUs, due to the design restrictions of web browsers, all contemporary mining libraries are limited in CPU-only mining.

\subsection{How does web mining work?}
The large growth of web-based cryptocurrency mining
started with the release of Coinhive's \js\ implementation 
of a Monero miner in  September 2017~\cite{coinhive-api}. This \js-based
miner, which computes hashes as a Proof-of-Work, could be easily 
included in any website for enabling publishers to utilize visiting 
users' CPUs as a way to monetize the visits to their websites. 

As a result, upon visiting a miner-supported website, the user
along with the website's necessary content (i.e., HTML, \js\, CSS) 
receives a mining library, too. Usually these mining libraries are 
provided by third parties, which we will refer to as Mining Service 
Providers (\controller), who are responsible for maintaining the source 
code, controlling the synchronization of computations, collecting 
the computed hashes and sharing the profits with the publishers.

Upon rendering, a miner establishes a persistent connection using HTML5's WebSocket 
API~\cite{websocket} with a remote third-party server, which is typically operated by the
\controller\ (e.g., coinhive.com), in order to communicate with the 
service/mining pool. Through this channel the miner receives 
periodically proof-of-work tasks and reports 
the successfully computed hashes.  

Coinhive (like similar projects e.g., CryptoLoot, JSEcoin)
leverage the capabilities of modern web browsers
and the advances of \js\ in order to make efficient mining 
through parallelization. Hence, these mining libraries typically 
use multiple threads on the visitor's system, through the 
utilization of HTML5's Web Workers API~\cite{web-workers}.
Furthermore, apart from performing computations in multiple threads, 
the use of web workers enables background processing.
This way, the heavy mining process running on the background 
does not intervene with the core functionality of the website, 
which can be rendered unhampered in the user's display.
In addition, typically mining libraries include code written 
in web assembly~\cite{webassembly}, for running all the 
necessary complex mathematic functions nearly as fast 
as native machine code.

\subsection{Cryptojacking}
Of course the increasing growth of web-based miners does not create opportunities only for legitimate publishers, but cyber-attackers as well.
Soon after the release of the first mining library from Coinhive in 2017, numerous incidents have been reported~\cite{bad-packets-report} of attackers 
injecting mining code snippets in websites with increased audience. As a result, the browsers of unaware visitors are forced to mine cryptocoins, thus abusing the
system's resources. This so-called Drive-by Mining or \emph{cryptojacking} takes place either by compromising embedded third party libraries or by delivering malicious mining code
through the ad ecosystem~\cite{ad_mining}. For example, the compromisation of a single screen reader third party (i.e., Browsealoud~\cite{browselowd}) resulted 
in infecting more than 4000 websites that were using it.
Victims of cryptojacking have been popular and prestigious websites like the official webpage of Cristiano Ronaldo~\cite{ronaldo}, the websites of CBS Showtime~\cite{showtime} and PolitiFact~\cite{politifact}, and the UK's Information Commissioner's Office~\cite{commissioner}, Student Loans Company and National Health Service websites~\cite{ukJacked}.

Of course, the notion of cryptojacking does not only include compromised websites but also 
websites that use web mining as a method for monetization but abstain from informing the users about the existence of cryptominers.
Indeed, contrary to the digital advertising were visitors can identify (in most of the cases~\cite{adlabels}) the ad-impressions, 
in web mining it is not easy for the visitors to perceive the existence of an included miner.
Consequently, web cryptojacking is a malicious action that abuses the user's processing power and
includes any \webmining\ attempt without the user's consent irrespectively whether the mining code has 
been legitimately deployed by the publisher of the website or a malicious actor that hijacked the website.

All the recently reported cases of cryptojacking, inevitably contributed towards giving a bad reputation to a possibly viable monetization method for web publishers.
As a consequence, we see a mine blocking movement rising, similar to the ad-blocking trend, where blocking extensions following an all-out model, block \emph{any} detected 
mining library even if this requests the permission of the user in order to start 
mining.

%% file: dataset.tex
\begin{figure}
	\centering
	\includegraphics[width=1\columnwidth]{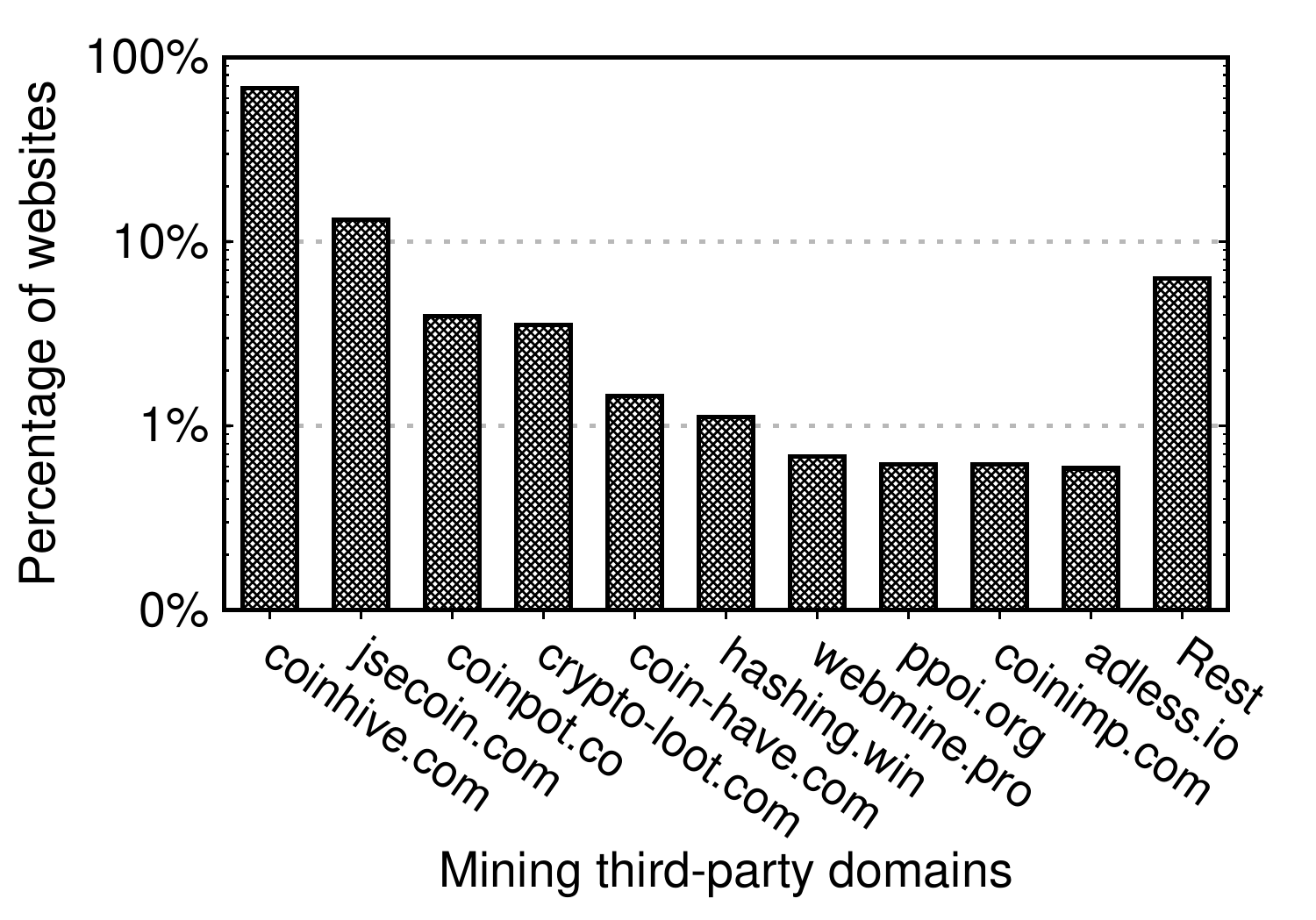}
	\caption{Cryptomining market share per third party library in our dataset. As seen, Coinhive owns the dominant share (69\%) when JSEcoin 
		follows with 13\%.}
	\label{fig:popularity}
\end{figure}

\section{Data collection and analysis}
\label{sec:dataset}

In order to gather the necessary data for our study, we first collect several coin-blocking blacklists~\cite{coinblockerlists} including the ones
used by the 5 most popular mine-blocking browser plugins: Coin-Blocker~\cite{coinBlocker}, No Mining~\cite{nomining}, MinerBlock~\cite{minerblock}, noMiner~\cite{nominer} and 
CoinBlock~\cite{coinblock}. By merging these blacklists we compose a list 
that contain 3610 unique entries of mining libraries and keywords. 
Then we use each of these entries to query PublicWWW's~\cite{publicWWW} 
dataset of the top 3 million pre-crawled landing pages, and we find 
107511 mine-including domains. 
It should be noted that the domains we collected that perform mining, are 
ranked in the range from 1353 to 960540 in the Alexa rank of popular 
websites, and that the majority of them are based in the USA, Russia and Brazil.

The mining websites we collected, include more than 27 different 
third party miners, such as Coinhive\footnote{https://coinhive.com/}, CryptoLoot\footnote{https://crypto-loot.com/} and CoinHave\footnote{https://coin-have.com/}. In Figure~\ref{fig:popularity}, we
present the portion of websites in our dataset that use each one of these 
third party mining libraries. 
As can be seen, besides the large variety of mining libraries,
there is a monopolistic phenomenon in the market of cryptominers, 
with Coinhive owning the dominant share (69\%), when from the rest of 
its competitors only JSEcoin miner surpassing 10\%. Furthermore, our 
dataset includes 2 different cryptocoins such as Monero~\cite{monero} and JSECoin~\cite{jsecoin}.

Apart from these miner-supported websites, we also collected an equal 
number of ad-supported ones, which are among the same popularity 
ranking range. We process each of these domains and by using the 
blacklist of Ghostery open-source adblocker, we enumerated all 
ad-slots in the landing page. The average number of ad-slots per 
website was 3.4. Finally, Table~\ref{tbl:summary} summarizes the 
contents of our dataset.

\subsection{\testbed\ framework for utilization analysis}
\begin{figure}[t]
	\centering
	\includegraphics[width=0.99\columnwidth]{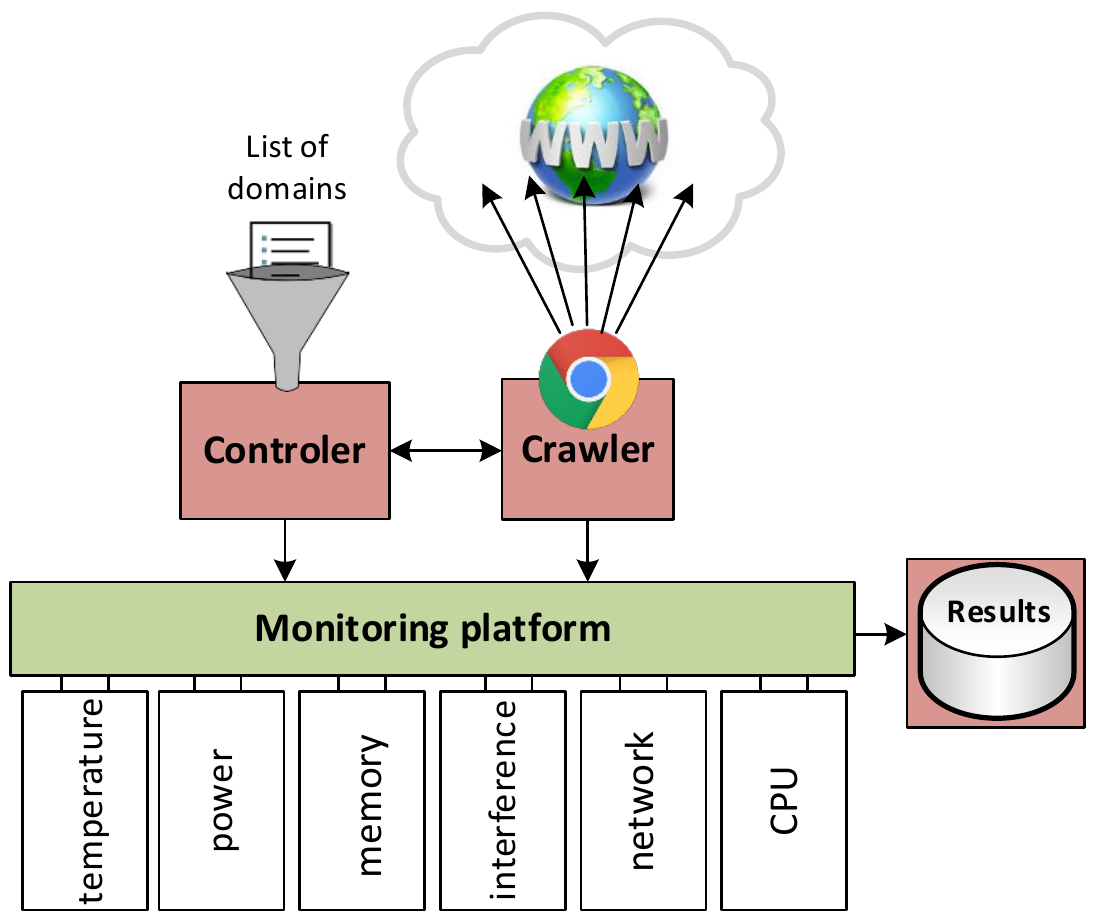}
	\caption{High level overview of our web measuring testbed. A Chrome-based platform fetches each website for a specific time when the different components measure its resources for behavioral analysis.}
	\label{fig:testbed}
\end{figure}

To measure the costs that each domain in our dataset imposes on the user, 
we designed and developed \testbed: a web measuring testbed.
A high-level overview of the architecture of \testbed\ is presented in Figure~\ref{fig:testbed}.
The \testbed\ framework follows an extensible modular design, and consists of several measuring components that work in a plug-and-play manner. Each such plug-in component is able to monitor usage patterns in different system resources (e.g., memory, CPU, etc.). The main components of our platform, as can be seen in Figure~\ref{fig:testbed}, currently include:

\begin{table}[t]
\centering
	\begin{tabular}{l|c}
	{\bf Type} & {\bf Amount} \\ \hline
	Blacklist entries & 3610 \\
	Miner-supported websites & 107511 \\ 
	Ad-supported websites & 100000 \\ 
	Unique third-party miners & 27 \\
	Crypto-coins & Monero, JSEcoin \\ \bottomrule
	\end{tabular}
	\caption{Summary of our dataset}
	\label{tbl:summary}
\end{table}

\begin{enumerate}[label=\Alph*.]
\item {\bf crawler} component, which runs the browser (i.e., Google Chrome) in a headless mode. The crawling is responsible of stopping and purging \emph{any} state after a website probe (e.g., set cookies, cache, local storage, registered service workers, etc.), and listening to the commands of the main controller (i.e., next visiting website, time interval, etc.).
\item {\bf main controller}, which takes as input a list of domains and the visiting time per website. It is responsible of scheduling the execution of the monitoring components and the results printer.
\item {\bf monitoring platform}, which is responsible for the per time interval execution of the monitoring modules. This platform was build in order to be easily expandable in case of future additional modules. 
\item  {\bf results printer}, that dumps in a local database the monitoring results.
\end{enumerate}

For the scope of this analysis, we developed 6 different modules to measure the utilization that miners perform in 6 different system resources: 
\begin{enumerate}
	\item memory activity (physical and virtual), by using the psrecord utility~\cite{psrecord} and attaching to the crawling browser tab's pid.
	\item CPU utilization per core, by using the dedicated linux tool process status (ps~\cite{ps}).
	\item system temperature (overall and per core), by leveraging the Linux monitoring sensors ({\tt lm\_sensors}\cite{lmsensors}).
	\item network traffic, by capturing (i) the network packets through {\tt tcpdump} and (ii) the HTTP requests in the application layer along with their metadata (e.g., timing, initiator, transferred bytes, type, status code), in pcap and HAR files respectively.
	\item process interference, to infer the degradation of user experience caused by the heavy CPU utilization of mining processes. Specifically, this module consists of a CPU intensive benchmarking that includes multi-threaded MD5 hash calculations.  
	\item energy consumption, 
	by installing in our machine an external Phidget21 power sensing package~\cite{phidget2,phidget1}. Phidget enable us to accurately measure the energy consumption of the 3 ATX power-supply lines  (+12.0a,  +12.0b  +5.0,  +3.3  Volts). The  12.0  Va
	line powers the processor, the 5.0V line powers the memory, and the 3.3V line powers  the  rest  of  the  peripherals  on  the  motherboard.
\end{enumerate}

The source code of \testbed\ along with the developed monitoring modules are provided open source\footnote{\testbed\ source: https://github.com/panpap/ webTestbench}.

\noindent{\bf Methodology:} In order to explore the different resource utilization patterns for miner- and ad- supported websites, we load our domain dataset in 
\testbed\ and we fetch each landing page for a certain amount of time. During this period the network monitoring module captures all outgoing HTTP(S) requests 
of the analyzed website. Additionally, the modules responsible for measuring the energy consumption, the CPU and memory utilization and the temperature report
 the sensor values in a per second interval. 
By the end of this first phase, \testbed\ erases any existing browser state and re-fetches the same site. This time, the 
only simultaneously running process is the interference measuring module which reports its progress at the end of the second phase.

%% file: measurements.tex
\section{Analysis}
\label{sec:measurements}
In this section, we aim to explore the profitability of the cryptomining web monetization model for the publishers, and to compare it with the current 
dominant monetization model of the web: digital advertising. Towards that direction, we assess the costs imposed on the user side in an attempt to 
determine the overheads a website's visitor sustains while visiting a miner-supported website. For the following experiments, we use a Linux 
desktop equipped with a Hyper-Threading Quad-core Intel I7-3770 
operating at 3.90 GHz, with 8 MB SmartCache, 8 GB RAM and an Intel 
82567 1GbE network interface.

\subsection{Profitability of publishers}
In the first set of experiments, we set out to explore the profitability of 
cryptominers and compare it to the current digital advertising model. 
Thereby, in the first experiment we simulate the monthly profit of the 
two strategies for a website of moderate popularity: 100,000 visitors/month. Studies~\cite{userTime} have measured the average duration of a website 
visit being around 1 minute. For this experiment, we use the popular 
Monero mining library of Coinhive which currently provides a rate of 
0.0001468/1M hashes. This means that the publisher gets 0.0001468 
Monero (XMR) (at the time of the experiment: 1 Monero=205 USD) per 1 million successfully calculated hashes. Apart from the visit duration, the amount 
of total calculated hashes of a publisher depends on the computation 
power of the visitors' devices. Thus, in this experiment, in order to cover 
a wider range of CPU hashrate capabilities~\cite{hashratesMonero}, we 
use 4 different levels of computation rates (Dual-core (3.50GHz) with Hyper-Threading: 50 Hash/sec, Quad-core (3.30GHz): 100 Hash/sec, 
Quad-core (4.00GHz) with 
Hyper-Threading: 200 Hash/sec and Octa-core (4.3GHz) 300 Hash/sec).

\begin{figure}[t]
	\centering
	\includegraphics[width=1\columnwidth]{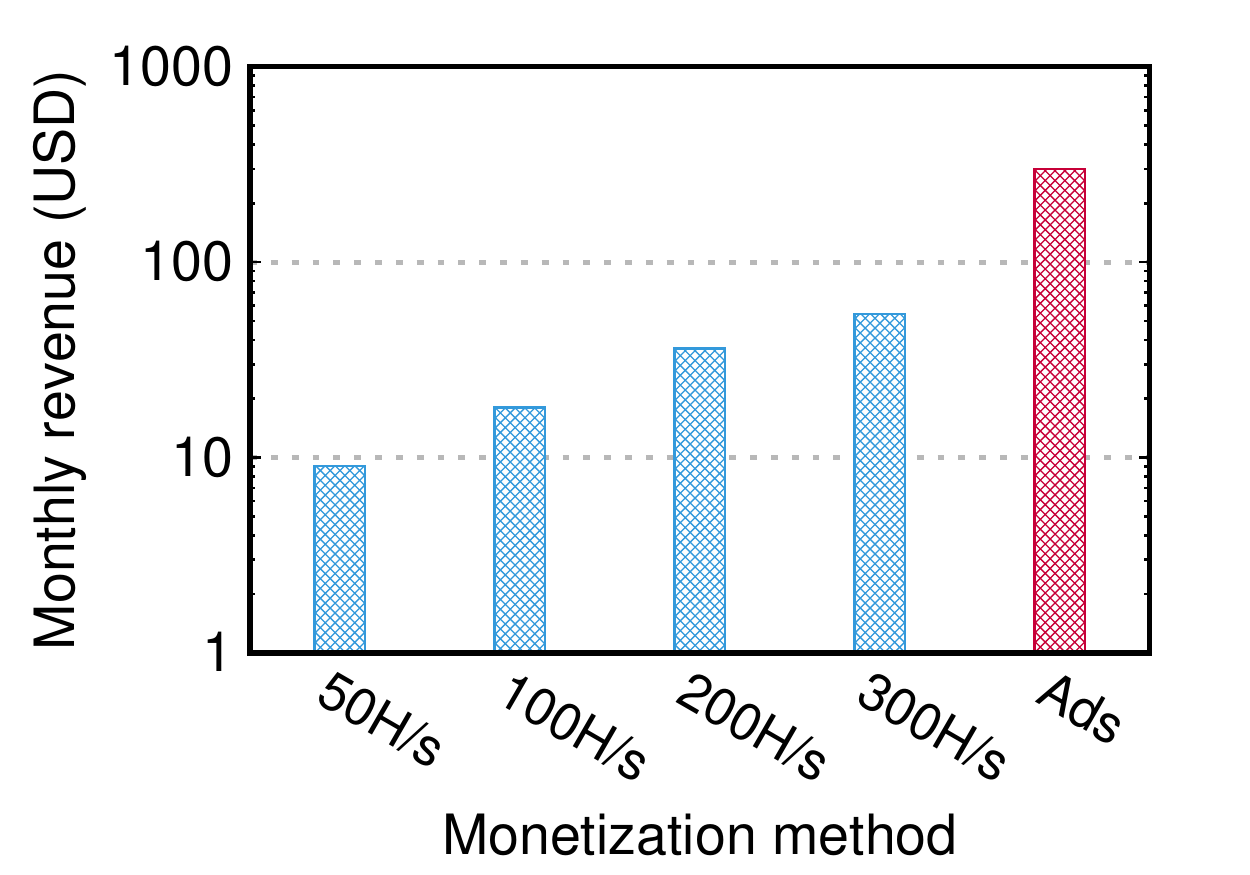}
	\caption{Estimation of monthly profit for the different monetization methods for a website with 100K visitors and average visit duration of 1 minute. Even for visitors with powerful devices (300Hashes/sec), a publisher gains $5.5\times$ more revenue by including 3 ads in its website.}
	\label{fig:profitPermonth}
\end{figure}

Apart from the profit from cryptomining, in this experiment we also compute the monthly revenue of the same website in the case of following the traditional advertising model. The most popular medium for personalized ad-buying nowadays~\cite{programmaticAd,programmaticAd2} is the 
programmatic instantaneous auctions. In this model, advertisers bid in real time auctions for each available ad-slot of a publisher's inventory 
based on how well the visitor's interests match their advertised product. As described in Section~\ref{sec:dataset}, the average number of 
ad-slots in an ad-supported website is 3 and the median
charge price per ad impression as measured in previous studies~\cite{imcRTB} is 1 CPM. As can be seen in Figure~\ref{fig:profitPermonth}, for the average duration of a user's visit, the publisher even when achieving an average computation rate from visitors of as high as 300Hash/sec, gains 5.5x 
more revenue when using ads instead of cryptomining.
Our simulation results are verified by the real world experiment of M. Cornet~\cite{3dayExperiment}.

\begin{figure}[t]
	\centering
	\includegraphics[width=1\columnwidth]{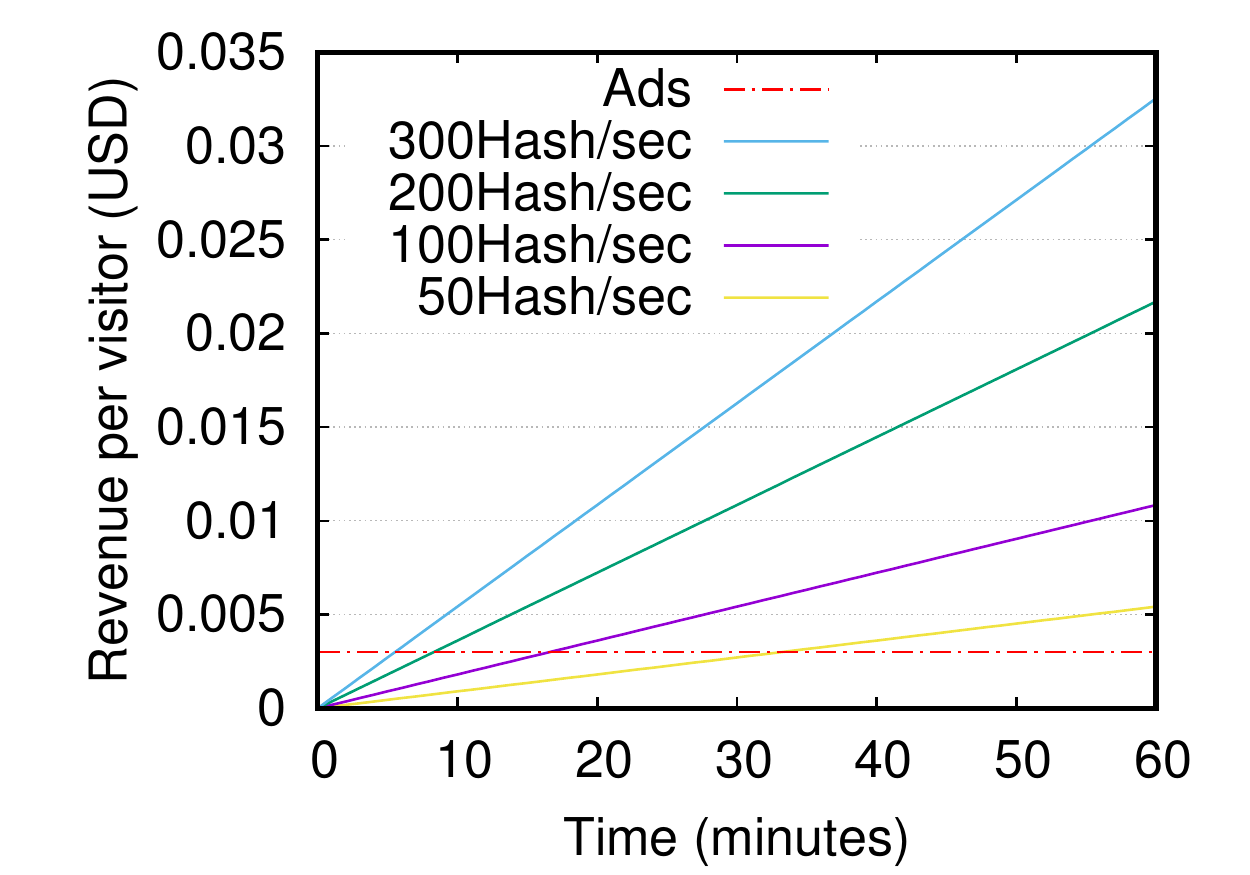}
	\caption{Revenue per visitor for a website running in a background tab. In order for a publisher to gain higher profit from mining than using ads (3 ad-slots), a visitor  must keep his tab open for duration $>5.3$ minutes (depending on the their device).}
	\label{fig:profitPermonth2}
\end{figure}

It is apparent thus, that even with visitors equipped with powerful devices, \emph{time matters for a miner-supported website}. 
Indeed, recent studies~\cite{mineFever} show that the majority of miner-supported websites provide content that can keep the visitor 
on the website for a long time. Such content includes TV, video or movie streaming, flash games, etc. Of course in cryptomining, the user 
does not need to interact with the website's content per se. 
As a consequence, there are numerous deceiving methods (e.g., pop-unders~\cite{popunder}) currently in use, aiming to allow the 
embedded miner to work in the background for as long as possible.

In the next experiment, we set out to identify the minimum time the publisher's website needs to remain open in the background in a visitor's browser 
tab in order to gain profit higher than when using ads. In Figure~\ref{fig:profitPermonth2} we simulate the revenue per visitor for a website 
running in the background and we use the same hash-rate levels as above. As shown, the miner-using publisher, in order to produce revenues higher 
than when ads are delivered, must keep its website open in a user's browser for duration longer than 5.3 minutes in the case of a user equipped with a high performance device (300Hash/sec), or longer than 33,1 minutes in case of a low performance device of 50Hash/sec!

\subsection{Costs imposed on the user side}
After estimating the revenues of a publisher for the different monetization methods, it is time to measure the costs each of this method imposes 
on the user.

\subsubsection{CPU and Memory Utilization}
In the first set of experiments, we explore the average CPU and memory utilization of mining supported websites.
Note at this point, that the intense of mining is tunable. The majority of mining libraries allow the publishers to fine tune the number of threads and the throttling of their included miner. In this experiment we fetch each website in our two subsets for 3 minutes using \testbed\ and we extract the distribution of its CPU utilization 
through time. In Table~\ref{tbl:cpuLoad} we report the average values for the median, the 10th and 90th percentiles. As we see, the median miner-supported website utilizes the visitor's CPU up to 59 times more than an ad-supported website.

\begin{table}[b]
	{ 
		\begin{tabular}{l||c|c|c}
			{\bf Type} & \textbf{ $10^{th}$ Perc.} & \textbf{Median} &  \textbf{$90^{th}$ Perc.}  \\ \hline
			Advertising & 3.33\% & 9.71\% & 17.19\%  \\ 
			Mining	& 560.11\% & 574.01\% & 580.71\% \\ \bottomrule
		\end{tabular}
		\caption{Distribution of the average CPU Utilization for the different monetization methods. The median miner-supported website utilizes 59x more the user's CPU than the median  ad-supported website.}
		\label{tbl:cpuLoad}
	}
\end{table}

\begin{figure}[t]
	\centering\vspace{-0.55cm}
	\includegraphics[width=1.07\columnwidth]{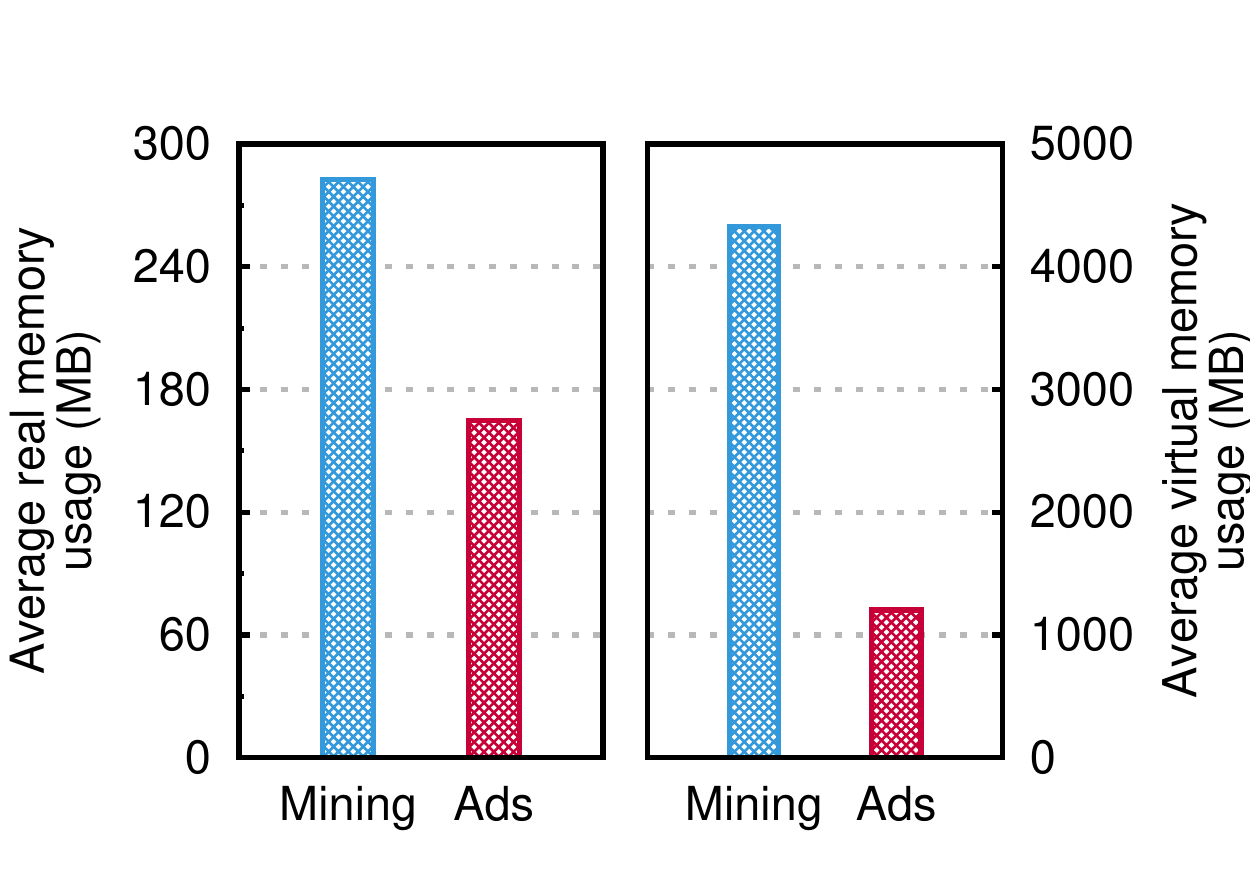}\vspace{0.1cm}
	\caption{Distribution of average real and virtual memory utilization through time. Miner-supported websites although 
	reserve (3.59x) larger chunks of virtual memory, require 1.7x more MBytes of real memory than ad-supported websites.}
	\label{fig:memory}
\end{figure}

In the same way, we measure the utilization of the visitors main memory and in Figure~\ref{fig:memory} we plot the average values for both real and virtual memory activity.
As expected, miners do not utilize memory as heavy as CPU. In particular, we see that on average the miner-supported websites require 1.7x more 
space in real memory than the ad-supported websites.

\begin{table*}[t]
	\centering
	\begin{tabular}{c|c||c|c|c}
		\bf Component & {\bf Type} & {\bf $10^{th}$ Percentile} & {\bf Median} &  {\bf $90^{th}$ Percentile}  \\ \hline
		\multirow{2}{*}{CPU \& Network adapter} & Advertising & 31.88 Watt & 32.39 Watt& 34.17 Watt  \\ 
		& Mining	& 63.35 Watt & 67.60 Watt & 71.22 Watt \\ \hline
		\multirow{2}{*}{Main Memory} & Advertising & 4.37 Watt & 4.46 Watt & 5.35 Watt \\
		& Mining & 4.76 Watt & 4.99 Watt & 5.67 Watt \\
		\bottomrule
	\end{tabular}
	\caption{Distribution of the average consumption of power for the different monetization methods. 
		The median miner-supported website forces the user's device to consume more power than the median ad-supported website: 2.08x  and 1.14x more power for the CPU and the memory component, respectively.}
	\label{tbl:power}
\end{table*}

\begin{figure}[t]
	\centering
	\includegraphics[width=1\columnwidth]{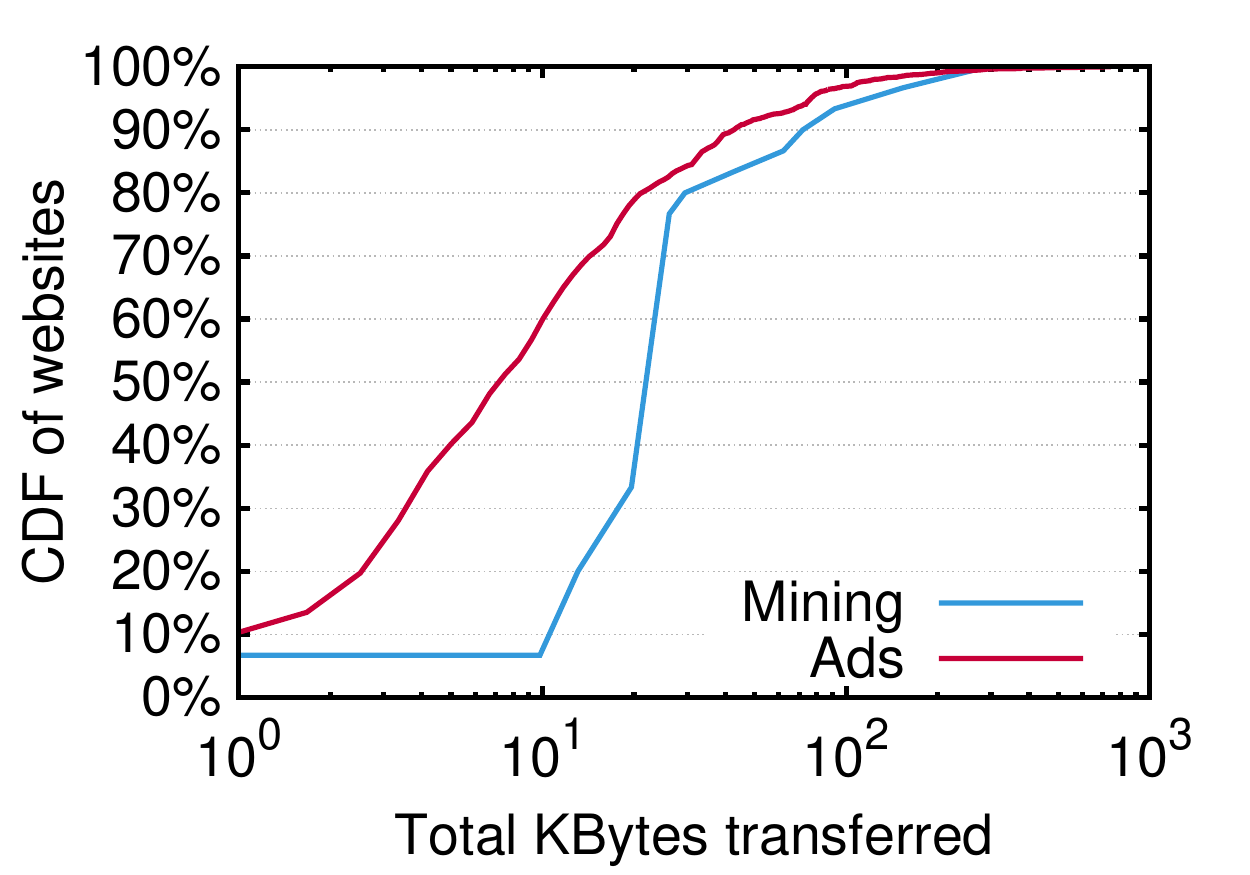}
	\caption{Distribution of the total transmitted volume of bytes per website for a visit duration of 3 minutes. 
		The median miner-generated traffic volume is 3.4x larger than the median ad-generated. In 20\% of the websites the difference reduces significantly (less than 2x).}
	\label{fig:networkVs}
\end{figure}

\subsubsection{Network Activity}
Next, we measure the network utilization of the average mining-supported 
website. As discussed in Section~\ref{sec:background}, a mining library 
needs to periodically communicate with a remote third party server (i.e., the \controller's server) in order to 
report the calculated hashes but also to obtain the next PoW. This 
communication in the vast majority of the libraries in our dataset takes place 
through a special persistent channel that allows bidirectional communication. To assess 
the network activity of web miners, we use the network capturing module of \testbed\ and we 
monitor the traffic of each (ad- and miner-supported) website for 3 minutes.

Based on the detected third-party mining library, we isolate the web socket communication between its in-browser 
mining module and the remote \controller\ server. In order to compare this PoW-related communication of miners with the corresponding ad-related 
traffic of ad-supported websites, we utilize the open-source blacklist of the Disconnect browser extension\footnote{Disconnect: https://disconnect.me/}
to isolate all advertising related content. In Figure~\ref{fig:networkVs}, we plot the distribution of the total transmitted volume of bytes 
per website for the visit duration of 3 minutes. Although the web socket communication of miners consists of small packets of 186 Bytes on average,
we see that in total the median PoW-related communication of miner-supported websites transmitted 22.8 KBytes, when the median ad-traffic volume of 
ad-supported websites was 6.7 KBytes. This means that the median miner-generated traffic volume is 3.4x larger than the median ad-generated.
In this experiment, we see that the network utilization patterns depend not only on the throttling of the miner but also on the different implementations. 
For example, while using the same portion of CPU, the miner of coinhive.com transmits on average 0.6 packets/sec, webmine.cz: 2.2 packets/sec, 
cryptoloot.com: 4.7 packets/sec and jsecoin.com: 1.3 packets/sec.

\begin{figure}[t]
	\centering
	\includegraphics[width=1\columnwidth]{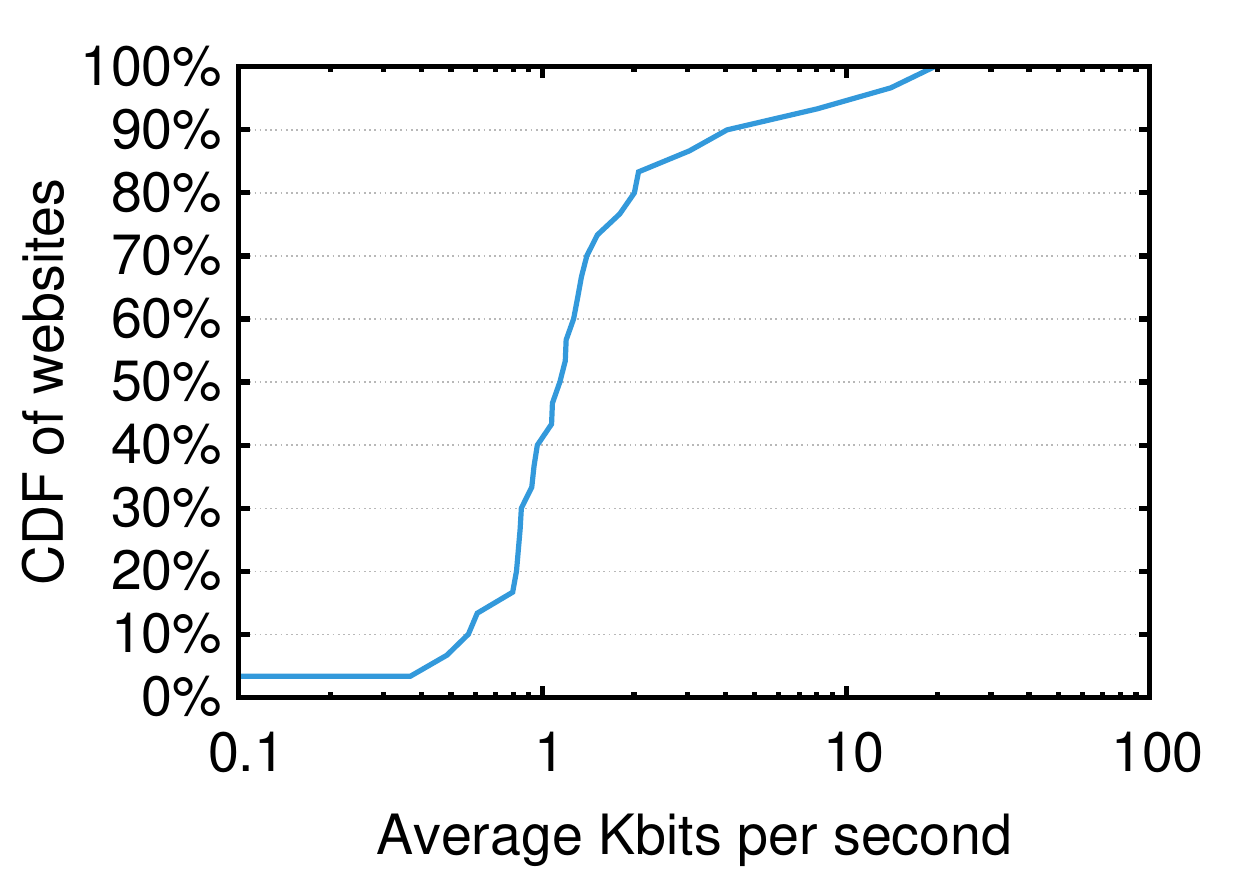}
	\caption{Distribution of the transmitted bit rate per miner-supported website in our dataset. The median in-browser miner 
		communicates with its remote \controller\ by transmitting 1.168 bits per second.}
	\label{fig:network}
\end{figure}

In Figure~\ref{fig:network} we plot the distribution of the average data transfer 
rate per miner-supported website in our dataset. As shown, the median  communication between the miner and the \controller\ has a transfer rate of 
1 Kbit per second (or 146 Bytes/sec). As in the previous experiment, the rate highly depends on the mining library, 
with some of them reaching up to 14 Kbit per second. At this point, recall that the PoW-related communication between the in-browser 
miner and the \controller\ holds for as long as the miner is running, and 
as we saw in Figure~\ref{fig:profitPermonth2} a miner must run for longer 
than 5.3 minutes in order to produce revenues higher than ads. This means 
that for the median case, the total volume of bytes transferred will exceed 46 KBytes. 

In the case of a user that browses through
a cellular (4G) network\footnote{Considering the average prices per byte in USA and Europe~\cite{dataplans2,dataplan3,dataplans}}, 
the monetary cost imposed is 0.000219\$ per minute on average, while browsing a miner-supported website. On the other hand, a publisher
including a coinhive miner in its website from the same user earns 0.000409\$ per minute (considering that the user provides an average hash rate of 227Hash/sec 
as in~\cite{3dayExperiment}). Hence, we see that cellular users, among other costs while visiting miner-related websites, pay a monetization cost that is only 53\% 
less than the revenue of the publisher. 

\subsubsection{Power Efficiency}
Of course the intensive resource utilization of cryptominers 
affects also the power consumption of the visitor's device, which has a direct 
impact on its battery autonomy. In the next experiment, we measure the power consumed by (i) main 
memory and (ii) CPU and network adapter components of the user's device while visiting miner- and ad- supported websites for a 3 minute duration.
In Table~\ref{tbl:power}, we report the average median, 10th and 90th percentile values for all websites in our dataset.
As we can see, there is a slightly increased (1.14x more than ad-supported websites) consumption of the memory component 
in miner-supported websites. However, we see that the heavy computation load of cryptominers significantly increases 
the CPUs and network adapters consumption, making miner-supported websites consume 2.08x more energy than ad-supported websites!
This means that a laptop able to support 7 hours of consecutive traditional ad supported browsing, would support 
3.36 hours of mining-supported browsing.

\begin{figure}[t]
	\centering
	\includegraphics[width=1\columnwidth]{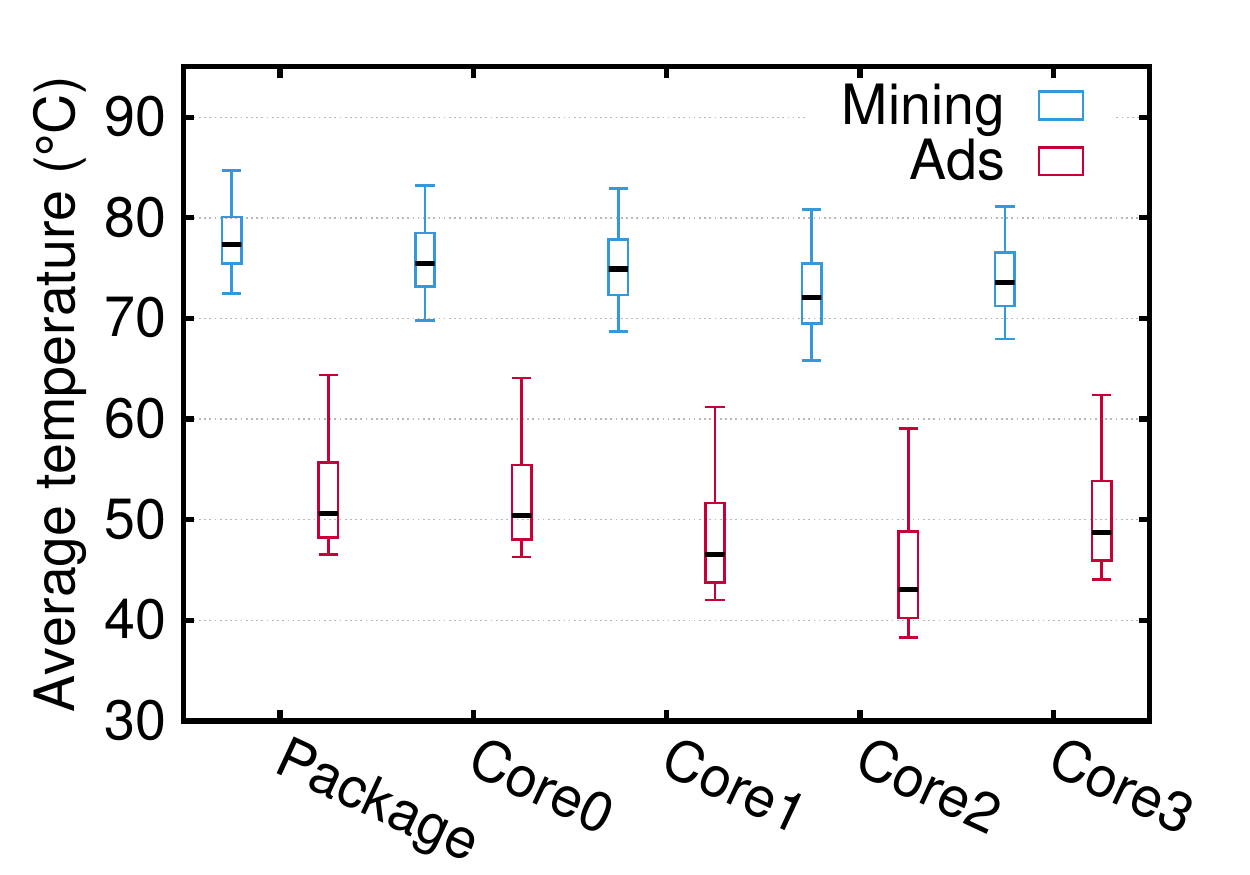}
	\caption{Distribution of average temperatures per system's core. When the visited website 
		includes miner, the average temperature of the cores may reach up to 52.8\% higher ($73-77^{\circ}$ Celsius) than when with ads.}
	\label{fig:temperature}
\end{figure}

\subsubsection{System Temperature} 
The increased electricity powering of the visitor's system results to an 
increased thermal radiation.
During the above experiment, we measure the distribution of the per-core temperatures while visiting each website in our dataset for 3 minutes.
In Figure~\ref{fig:temperature} we present the average results for the percentiles: 10th, 25th, 50th, 75th, 90th. As we can observe, the 
core temperatures for miner-supported websites are constantly above the optimal range of $45-60^{\circ}$ Celsius~\cite{optimalTemp2,optimalTemp}. In particular,
the visitor's system operates for most of the time in the range of $43-50^{\circ}$ Celsius while visiting ad-supported websites. When the visited website 
includes miner, the average temperature of the cores reaches up to 
52.8\% higher, in the range of $73-77^{\circ}$ Celsius, when in 10\% 
of the cases it may reach higher than $84^{\circ}$ Celsius.

To that end, with regards to the costs imposed to the user, 
high temperatures may lead to degraded system performance 
and poor user experience. 
Apart from that, constantly running a commodity device (such as a mobile phone, laptop or desktop PC) at high temperatures, without a proper cooling mechanism, may significantly decrease the hardware's lifespan in the long term or even cause physical damage by thermal expansion.

\subsubsection{Effects on Parallel Running Applications}

It is apparent that the heavy utilization of the visitor's CPU is capable of 
affecting the overall user's experience not only in the visited website, 
but in parallel processes and browser tabs, too. 
Indeed, for as long as the browser tab of a mining-supported 
website is open, the multi-threaded computations of the embedded miner 
leaves limited processing power for the rest of the running applications. 
To make matters worse, as part of a PC's own cooling system, the 
motherboard in case of increased temperatures may instruct the CPU 
component to slow down (in the case of overheating, motherboard 
may force the whole system to turn off without warning)~\cite{overheating}.

\begin{figure}[t]
	\centering
	\includegraphics[width=1\columnwidth]{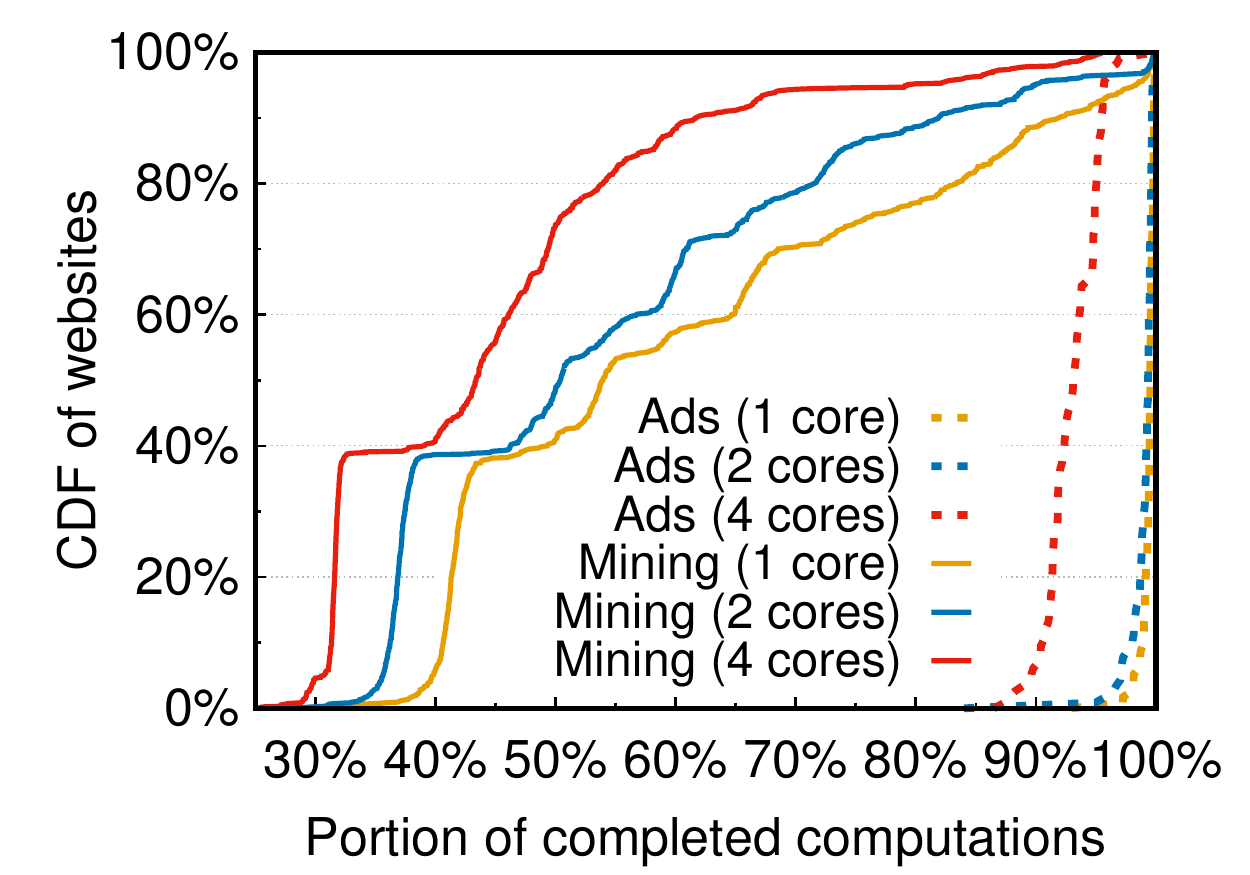}
	\caption{Impact of background running miner- and ad-supported websites to a user's process. When the majority of ad-supported websites have negligible 
	effect in other processes, the median embedded miner in our dataset through its heavy CPU utilization may cause a performance degradation of higher 
	than 46\% to a parallel running process.}
	\label{fig:interference}
\end{figure}

To assess how these factors may affect parallel running processes in the visitor's device, in the next experiment, 
we use the interference measuring module of \testbed\ and we measure the performance overhead caused by background running miners.
This module, introduces computation workloads to the system to emulate a parallel running process of the user.
Specifically, \testbed\ fetches each website in our dataset for the average visit duration (i.e., 1 minute), in parallel conducts multi-threaded MD5 hash calculations, and in the end reports the number of successful calculated hashes.
In order to test the performance of the user's parallel process in different computation intensity levels, we visit each website using 3 setups 
for the MD5 process, utilizing in parallel 1, 2, and 4 cores of the CPU.
In addition, we run the MD5 process alone for 1 minute to measure 
the maximum completed operations.

In Figure~\ref{fig:interference}, we plot the distribution of the portion of completed operations per website.
As expected, we can clearly observe that when there is a miner-supported website running in the user's browser, 
the performance of the user's processes that run in parallel is severely affected. In particular, we see that the median 
miner-supported website forces the parallel process (depending on its computation intensity) to run in 54\%, 50\% or even 43\%  
of its optimal performance, thus causing an overall performance degradation that ranges from 46\% to 57\%! Additionally, we see a 39\% of miners 
greedily utilizing all the system's CPU resources causing a performance reduction of 67\% to the parallel process.


Moreover, in this figure we plot the interference that ad-supported websites impose to a parallel process.
As expected, the impact is minimal and practically only processes with full CPU utilization are affected, facing 
a performance degradation of less than 10\% for the majority of such websites. 
This slight performance degradation is caused by the \js\ code responsible for ad serving, user tracking, analytics, etc., deprives scheduling time quantums from the parallel process.
%

It is of no doubt, that such severe performance degradation when the 
user is visiting a mining-supported website can cause glitches, or even 
crushes in other, parallel, CPU utilizing applications (like movie playback, video 
calling, file compression, video games, etc.), thus ravaging the user's 
experience. Of course, this performance degradation does not only affect parallel 
running applications of the user but also mining operations from other open browsing tabs.

Indeed, a miner can achieve full utilization when the user has visited the miner-supported $website1$.
However, when the user opens a second miner-supported $website2$ the maximum utilization for both, as well as the revenues for $publisher1$ and $publisher2$, drop to a half.
It is easy thus to anticipate, that {\bf the scalability of cryptomining is limited since the more websites rely on 
\webmining\ for funding, the less revenues will be for their publishers.}
While this monetization model has that apparent drawback, in digital advertising each ad-supported website is totally 
independent from any parallel open browser tabs.

%% file: discussion.tex
\section{Discussion}
%
\subsection{User awareness}
The lack of adequate policies and directives regarding the proper use of cryptomining,
has raised a big controversy regarding the lack of transparency in miner-supported websites.
Many miner-supported websites do not inform the user about 
the existence of a miner, neither ask for the visitor's 
consent to utilize their system's resources for cryptocurrency mining. 

In one of the first law cases about web-based cryptocurrency mining, 
the Attorney General John J. Hoffman 
stated that ``\textit{no website should tap into a person's computer 
processing power without clearly notifying the person and giving them 
the chance to opt out}''~\cite{attorney}. 
As a consequence, whenever a user visits a website and she is not aware about 
the background \webmining\ procedure, irrespectively whether the mining code 
has been legitimately deployed by the publisher or a 
malicious actor that hijacked the website, this is considered as a
\textit{cryptojacking} attempt.
%

\subsection{Web-miner detection}
The emergence of \webmining, and especially the many reported 
cases of cryptojacking, pushed towards considering \webmining\ by 
default as a malicious operation. To that regard, many major Antivirus 
vendors consider \webmining\ as potential malware, and recently 
launched software products~\cite{malwarebytes, kaspersky} and browser extensions, 
such as NoCoin and minerBlock~\cite{blocking-extensions} for detecting and blocking in-browser 
miners. The vast majority of these approaches 
are mainly based on detecting outgoing requests to third parties that provide \js\ mining APIs (e.g., Coinhive) and mining pools. 

Moreover, most major browser vendors have started incorporating 
detections mechanisms to prevent the \webmining.  Opera has incorporated 
NoCoin in its built-in ad blocker mechanism to detect and block 
mining scripts~\cite{opera-nocoin}, and recently also incorporated 
this feature to the mobile versions of its browsers~\cite{opera-mobile}.
Chrome and Firefox are moving towards that direction, with the 
former removing from the Chrome Web Store all extensions that 
perform mining~\cite{chrome-extensions} and implementing a 
throttling mechanism that will limit CPU utilization for \js\ code 
running in the background~\cite{chrome-throttling}. Similarly, 
Firefox is investigating mechanisms for throttling background 
\js\ execution~\cite{firefox-throttling}.

However, even though these approaches currently seem to 
manage reducing the extend of \webmining\ and cryptojacking, 
they are not very robust against determined publishers/attackers, 
especially the ones based on the use of blacklists for detecting 
domains associated to miners. Recently we have seen miners
that try to avoid detection by only utilizing a percentage of the 
users' CPU processing power and by employing cloud-based 
proxy servers to handle all communication with the \controller's server~\cite{evading-blacklists}. Also, in many cases the 
mining code is highly obfuscated~\cite{obfuscation} to prevent 
pattern matching 
tools from detecting snippets of such suspicious code.  

\subsection{User consent}
In order to avoid criticism, and also avoid being blocked by 
Antivirus and mining blocking tools, Coinhive, which is 
promoted as a service that enables publishers to monetize 
their websites in a legitimate and clean way, provided 
the``AuthedMine''  \js\ mining implementation. This 
new mining service requires user's explicit consent to 
start running mining while visiting the 
website~\cite{authedmine}. Essentially, this approach shows
an attempt towards fighting cryptojacking and ``legitimizing'' 
Coinhive's mining service, as it prevents in-browser mining 
without the user's knowledge and consent.

Even though this approach is a step towards the right direction, 
it does not solve the problem of cryptojacking in its entirety. 
New mechanisms need to be designed and implemented by the
browser vendors for detecting the existence of mining scripts,
even if they are obfuscated, and informing the users about them. 
The provision of such efficient mechanisms, and the willingness
of publishers to only adopt legitimate mining services that inform
the user and require her consent for performing mining, can signal 
the emergence of a new monetization paradigm in the web.   

\subsection{Letting the users choose}
Since both digital advertising and \webmining\ impose a hidden 
cost on the user, each one in a different way, a new paradigm 
could be to inform the user about these costs in each case and 
give them the option to choose which of the two monetization schemes is 
more suitable for them. In the case of advertising the main cost 
to the user is related to the lack of privacy, while the cost of 
web-based mining is associated with higher energy consumption
(and battery drainage, device overheating etc). It seems that a 
viable option 
for web publishers would be to inform the users about these 
costs, and provide two different versions of their website
(i.e., one that serve ads and one that uses cryptoming), thus 
allowing the user to choose between the two schemes.


%% file: related.tex
\section{Related Work}
\label{sec:related} 
The advances of \js, which provide developers with parallel execution of their operations, and the 
development of more lightweight altcoins like Monero and Litecoin, enabled browser-based miners to grow. As a result,
content providers can deploy miner-supported websites without affecting the user experience.

Eskandari et al.~\cite{cryptojackingPaper}, in one of the first web mining related studies, analyzes the existing in-browser mining approaches and their profitability. In particular, the authors measured the growth of cryptomining during the last 
years by looking for mining libraries in Internet archive services. In addition, they collected a set 33K websites by querying 
for popular mining projects the Censys.io BigQuery dataset, and they studied the CPU utilization of the included miners.

AdGuard Research, which produces an ad-blocking software, in~\cite{mineFever} analyze the Alexa top 100,000 websites for cryptocurrency mining 
scripts in an attempt to measure the adoption of cryptominers in the web. The analysis revealed 220 of these websites using crypto-mining scripts with their aggregated audience being around 500 million people. The content of these hosting websites were usually content that could keep the user on the website for long and specifically movie/video/tv streaming (22.27\%), file sharing (17.73\%), Adult (10\%) and News \& Media (7.73\%) with the majority of them based in the U.S., India, Russia, and Brazil.

This rapid growth of web miners along with the frenzy increase of the cryptocurrency values, caused a serious debate over the Internet
regarding the ability of cryptomining to become an alternative to the current ad-supported model of Internet~\cite{debate,webminingwaste}.
In accordance with this debate, in this study, we compare the profitability of ad and cryptomining supported Internet services,
and we also measure the cost of cryptomining for the visitors. Of course, the advertising ecosystem also imposes costs on the user side.
Gui et al. in \cite{truthInAdvertising} measure the cost of mobile advertisements to the mobile application developer 
by performing an empirical analysis of 21 apps.
The authors consider several types of costs: (i) app performance, (ii) energy consumption, (iii) network usage, (iv) maintenance effort for ad-related code and (v) the users' feedback from app reviews.
Their results show that apps with ads consume on average 48\% more  CPU  time,  16\% more energy and 79\% more network data.
In addition, they found that the presence of ads in the apps affected the users' overall opinion leading to reduced ratings for the app. 

Papadopoulos et al. in~\cite{www18adcost}, analyze a year-long dataset with weblogs of volunteering users in an attempt to measure hidden 
costs of personalized advertising like the imposed monetary cost and privacy loss.
Authors compare their findings with the cost paid by the advertisers to deliver their impressions, and show that users pay $3\times$
more money to receive ads than what advertisers pay to deliver them. In addition, the authors used the leaked cookies from Cookie Synchronization 
to measure the anonymity loss of users. Results show that a handful of third parties can learn up to 10\% of the total unique userIDs of the median user across an entire year.

At the beginning, cryptomining was used mostly by shady websites that could not find proper revenue from digital 
advertising or illegally as a payload of malwares. Indeed, 
Wyke~\cite{sophos}, back in 2012, attempted to increase the awareness regarding the possibility of existing 
malwares delivering cryptomining payloads to infect user devices.
Botnets are examples of such malwares, which adopted mining to directly monetize the computational ability of a 
compromised computer. Huang et al.~\cite{botcoin} conducts a comprehensive study of existing Bitcoin mining malware, and presents the infrastructure and miner-bot orchestration mechanisms deployed in the wild.

The same advancements of contemporary browsers that boosted the growth of cryptomining as a model of web monetization, enabled also attackers
to perform cryptojacking. Recent reports from cybersecurity agencies~\cite{enisaCryptoJacking} aim to warn users about the 
emerging threat of cryptojacking. Indeed, there are numerous incidents already reported, where websites~\cite{showtime,ronaldo} got infected (either though malvertising~\cite{malvertising} or server compromisation)
with mining malware that abused visitors' devices. Dorsey, in~\cite{dorsey}, demonstrated his approach where by exploiting the ad ecosystem could 
widely deliver malware, which upon browser infection, could perform malicious computations like cryptomining on the user side.

%% file: summary.tex
\section{Summary And Conclusion}
Binded with the whopping values of cryptocoins, web-based cryptomining enjoys nowadays a steadily increasing adoption 
by service providers. More and more publishers choose \webmining for monetizing their websites in an attempt to abandon the sinking boat of digital advertising. But can cryptomining become a reliable alternative for the 
next day of the free Internet?

To respond to this exact question, in this paper we estimate the monthly revenue a publisher may gain by using cryptominers to monetize its content, and we compare
our results with the estimated revenue for the same publisher when using the traditional personalized advertising model. Then we compute the duration 
threshold for a website visit, after which a publisher can earn more revenue 
when using a cryptominer instead of ads.

After exploring the profitability for the side of the publisher, we measure the costs cryptominers impose on the side of the user.
Specifically, we analyzed the utilization patterns of miner-supported websites in the visitor's system resources like CPU, main memory and network.
Then, we study the impact of these utilization patterns (i) on the visitor's device by measuring the system's power consumption and temperature, and also (ii) on the 
visitor's experience while running other applications in parallel.

\subsection{Lessons Learned}
The findings of our analysis can be summarized as follows:
\begin{itemize}
\item for the average duration of a website visit, a publisher gains 5.5x more revenue by including 3 ad impressions in its website than by including a cryptominer.

\item to produce higher revenues with a miner than with ads, user must keep her browser tab open on the background for duration longer than 5.3 minutes.

\item the median miner-supported website utilizes up to 59x more the visitor's CPU and require 1.7x more space in real memory than ad-supported
websites.

\item the transfer rate of the median miner-\controller\ communication is 1 Kbit per second. For a user over cellular network the 
monetary cost imposed is on average 0.000219\$ per minute, when the publisher from the same user earns 0.000409\$ per minute.

\item the median miner-generated traffic volume is 3.4x larger than the corresponding ad-generated.

\item a visit to a miner-supported website consumes on average 2.08x 
more energy than a visit to an ad-supported website. 

\item a visitor's system while visiting a miner-supported website operates in up to 52.8\% higher temperatures than while visiting a website with ads.

\item web-miners severely affect parallel running processes. The median miner-supported website when running in the background may degrade 
even 57\% the performance of parallel running applications, thus wrecking 
the user experience.

\end{itemize}

\subsection{Can cryptomining become the next\\ monetization model for the web?}
After completing our analysis, we are now ready to respond to our motivating question regarding the ability of cryptomining to become the 
next primary monetization model for the web. Cryptomining can indeed 
constitute a reliable source of income for specific categories of publishers,
who can even increase their profits by providing content (movie/video streaming, flash games, etc.) that can keep the user on the website for a relatively longer time ($>5.3$ minutes). 

What is more, in these days, where EU regulators~\cite{gdpr} aim to reform the way user data are being collected and processed for targeted advertising, cryptomining provides a privacy-preserving monetization model that requires zero data from the users. 
However, in this study we see that the intensive resource utilization of current cryptomining libraries, imposes a significant cost on the user's device, accelerating the deterioration of its hardware. To make matters worse, this utilization also limits the scalability of cryptomining, 
since the more websites adopting miners the less portion of resources each of them will acquire from a user with multiple open tabs.  
To conclude, cryptomining indeed has the potential to become a reliable alternative for some content providers, but it is not capable of replacing the current ad-driven monetization model of the web.

%% file: paper.bbl
\begin{thebibliography}{100}

\bibitem{cpuonly}
Cpu coin list.
\newblock http://cpucoinlist.com/.

\bibitem{pujol}
Whotracks.me: Monitoring the online tracking landscape at scale.
\newblock https://arxiv.org/abs/1804.08959, 2018.

\bibitem{mineFever}
{AdGuard Research}.
\newblock Cryptocurrency mining affects over 500 million people. and they have
  no idea it is happening.
\newblock https://adguard.com/en/blog/crypto-mining-fever/, 2017.

\bibitem{malwarebytes}
P.~Arntz.
\newblock How to protect your computer from malicious cryptomining.
\newblock
  https://blog.malwarebytes.com/101/2018/02/how-to-protect-your-computer-from-malicious-cryptomining/,
  2018.

\bibitem{dataplans}
{AT\&T}.
\newblock Create your mobile share advantage plan.
\newblock https://www.att.com/shop/wireless/data-plans.html, 2018.

\bibitem{jsecoin}
J.~Banchini, J.~Sim, D.~Mallett, and T.~Howard.
\newblock Jsecoin is a cryptocurrency mined by webmasters and built for
  everyone.
\newblock Whitepaper, https://jsecoin.com/whitepaper.pdf.

\bibitem{attensionToken}
{BAT team}.
\newblock Basic attention token.
\newblock https://basicattentiontoken.org/.

\bibitem{overheating}
P.~Bates.
\newblock How heat affects your computer, and should you be worried?
\newblock
  https://www.makeuseof.com/tag/how-heat-affects-your-computer-and-should-you-be-worried/.

\bibitem{coinblock}
J.~Bechsen.
\newblock Coinblock.
\newblock https://addons.mozilla.org/en-US/firefox/addon/coinblock/.

\bibitem{minerblock}
I.~Belkacim.
\newblock Minerblock.
\newblock https://github.com/xd4rker/MinerBlock.

\bibitem{extortion}
J.~Bloomberg.
\newblock Ad blocking battle drives disruptive innovation.
\newblock https://www.forbes.com/sites/jasonbloomberg/
  2017/02/18/ad-blocking-battle-drives-disruptive-innovation/.

\bibitem{debate}
V.~Blue.
\newblock As online ads fail, sites mine cryptocurrency.
\newblock
  https://www.engadget.com/2017/12/15/as-online-ads-fail-sites-mine-cryptocurrency/,
  2017.

\bibitem{coinBlocker}
{Brandon-T}.
\newblock Coin-blocker.
\newblock https://github.com/Brandon-T/Coin-Blocker.

\bibitem{brave}
{Brave Software Inc. }.
\newblock Brave: A browser with your interests at heart.
\newblock https://brave.com/, 2018.

\bibitem{carrascosa2015always}
J.~M. Carrascosa, J.~Mikians, R.~Cuevas, V.~Erramilli, and N.~Laoutaris.
\newblock I always feel like somebody's watching me: measuring online
  behavioural advertising.
\newblock In {\em Proceedings of the 11th ACM Conference on Emerging Networking
  Experiments and Technologies}, page~13. ACM, 2015.

\bibitem{firefox-throttling}
C.~Cimpanu.
\newblock Firefox working on protection against in-browser cryptojacking
  scripts.
\newblock https://www.bleepingcomputer.com/news/
  software/firefox-working-on-protection-against-in-browser-cryptojacking-scripts/,
  2018.

\bibitem{chrome-throttling}
C.~Cimpanu.
\newblock Tweak to chrome performance will indirectly stifle cryptojacking
  scripts.
\newblock https://www.bleepingcomputer.com/news/security/
  tweak-to-chrome-performance-will-indirectly-stifle-cryptojacking-scripts/,
  2018.

\bibitem{qhostery}
{Cliqz GmbH}.
\newblock Ghostery makes the web cleaner, faster and safer!
\newblock https://www.ghostery.com/blog/.

\bibitem{cliqz}
{Cliqz GmbH}.
\newblock Cliqz: The no-compromise browser.
\newblock https://cliqz.com/en/, 2018.

\bibitem{coinhive-api}
Coinhive.
\newblock Monetize your business with your users' cpu power.
\newblock https://coinhive.com/\#javascript-api.

\bibitem{authedmine}
Coinhive.
\newblock A note to adblock and antivirus vendors.
\newblock https://authedmine.com/, 2018.

\bibitem{moneroHashRate}
{CoinWarz}.
\newblock Monero network hashrate chart and graph.
\newblock
  https://www.coinwarz.com/network-hashrate-charts/monero-network-hashrate-chart.

\bibitem{antiadblock}
D.~Coldewey.
\newblock Thousands of major sites are taking silent anti-ad-blocking measures.
\newblock
  https://techcrunch.com/2017/12/27/thousands-of-major-sites-are-taking-silent-anti-ad-blocking-measures/.

\bibitem{3dayExperiment}
M.~Cornet.
\newblock Coinhive review: Embeddable javascript crypto miner - 3 days in.
\newblock
  https://medium.com/@MaxenceCornet/coinhive-review-embeddable-javascript-crypto-miner-806f7024cde8,
  2017.

\bibitem{adblockingUsers}
M.~Cortland.
\newblock 2017 adblock report.
\newblock https://pagefair.com/blog/2017/adblockreport/.

\bibitem{hashratesMonero}
{CryptoMining24.net}.
\newblock Cpu for monero.
\newblock https://cryptomining24.net/cpu-for-monero/, 2017.

\bibitem{cryptonight}
{CryptoNote Tech}.
\newblock Cryptonote technology: Egalitarian proof of work.
\newblock https://cryptonote.org/inside.php\#equal-proof-of-work.

\bibitem{piratebay}
E.~V. der Sar.
\newblock The pirate bay website runs a cryptocurrency miner (updated).
\newblock
  {https://torrentfreak.com/the-pirate-bay-website-runs-a-cryptocurrency-miner-170916/}.

\bibitem{mining_trend}
D.~Desai, D.~Gandhi, M.~Sadique, and M.~Ghule.
\newblock Cryptomining is here to stay in the enterprise.
\newblock https://www.zscaler.com/blogs/research/
  cryptomining-here-stay-enterprise.

\bibitem{dorsey}
B.~Dorsey.
\newblock Browser as botnet, or the coming war on your web browser.
\newblock Radical Networks., 2018.

\bibitem{programmaticAd2}
{eMarketer Podcast}.
\newblock emarketer releases new us programmatic ad spending figures.
\newblock
  https://www.emarketer.com/Article/eMarketer-Releases-New-US-Programmatic-Ad-Spending-Figures/1016698,
  2017.

\bibitem{cryptojackingPaper}
S.~Eskandari, A.~Leoutsarakos, T.~Mursch, and J.~Clark.
\newblock A first look at browser-based cryptojacking.
\newblock In {\em Proceedings of IEEE Security \& Privacy on the Blockchain},
  S\&B 2018, 2018.

\bibitem{enisaCryptoJacking}
{European Union Agency for Network and Information Security (ENISA)}.
\newblock Cryptojacking - cryptomining in the browser.
\newblock
  https://www.enisa.europa.eu/publications/info-notes/cryptojacking-cryptomining-in-the-browser,
  2017.

\bibitem{dataplans2}
{FANDOM Lifestyle Community}.
\newblock Prepaid data sim card wiki - spain.
\newblock http://prepaid-data-sim-card.wikia.com/wiki/Spain, 2017.

\bibitem{ChromeBlocker}
K.~Finley.
\newblock Google's new ad blocker changed the web before it even switched on.
\newblock https://www.wired.com/story/google-chrome-ad-blocker-change-web/.

\bibitem{politifact}
B.~Fung.
\newblock Hackers have turned politifact’s website into a trap for your pc.
\newblock
  https://www.washingtonpost.com/news/the-switch/wp/2017/10/13/hackers-have-turned-politifacts-website-into-a-trap-for-your-pc/,
  2017.

\bibitem{followTheMoney}
P.~Gill, V.~Erramilli, A.~Chaintreau, B.~Krishnamurthy, K.~Papagiannaki, and
  P.~Rodriguez.
\newblock Follow the money: Understanding economics of online aggregation and
  advertising.
\newblock In {\em Proceedings of the 2013 Conference on Internet Measurement
  Conference}, IMC '13, 2013.

\bibitem{moneroGrowth}
{Global Coin Report}.
\newblock Here's how monero (xmr) gets to \$1,000.
\newblock https://globalcoinreport.com/heres-monero-xmr-gets-1000/, 2018.

\bibitem{Goldstein:2013:CAA:2488388.2488429}
D.~G. Goldstein, R.~P. McAfee, and S.~Suri.
\newblock The cost of annoying ads.
\newblock In {\em Proceedings of the 22Nd International Conference on World
  Wide Web}, WWW '13, pages 459--470, New York, NY, USA, 2013. ACM.

\bibitem{ad_mining}
D.~Goodin.
\newblock Ad network uses advanced malware technique to conceal cpu-draining
  mining ads.
\newblock
  https://arstechnica.com/information-technology/2018/02/ad-network-uses-advanced-malware-technique-to-conceal-cpu-draining-mining-ads/.

\bibitem{ukJacked}
P.~Greenfield.
\newblock Government websites hit by cryptocurrency mining malware.
\newblock https://www.theguardian.com/technology/2018/
  feb/11/government-websites-hit-by-cryptocurrency-mining-malware, 2018.

\bibitem{truthInAdvertising}
J.~Gui, S.~Mcilroy, M.~Nagappan, and W.~G.~J. Halfond.
\newblock Truth in advertising: The hidden cost of mobile ads for software
  developers.
\newblock In {\em Proceedings of the 37th International Conference on Software
  Engineering - Volume 1}, ICSE '15, pages 100--110, Piscataway, NJ, USA, 2015.
  IEEE Press.

\bibitem{botcoin}
D.~Y. Huang, H.~Dharmdasani, S.~Meiklejohn, V.~Dave, C.~Grier, D.~McCoy,
  S.~Savage, N.~Weaver, A.~C. Snoeren, and K.~Levchenko.
\newblock Botcoin: Monetizing stolen cycles.
\newblock In {\em Proceedings of Annual NDSS}, 2014.

\bibitem{altcoins}
{investing.com}.
\newblock All cryptocurrencies.
\newblock https://www.investing.com/crypto/currencies, 2018.

\bibitem{iqbal2017ad}
U.~Iqbal, Z.~Shafiq, and Z.~Qian.
\newblock The ad wars: retrospective measurement and analysis of anti-adblock
  filter lists.
\newblock In {\em Proceedings of the 2017 Internet Measurement Conference},
  pages 171--183. ACM, 2017.

\bibitem{attorney}
S.~C.~L. John~Hoffman, Jeffrey S.~Jacobson.
\newblock New jersey division of consumer affairs obtains settlement with
  developer of bitcoin-mining software found to have accessed new jersey
  computers without users’ knowledge or consent.
\newblock http://nj.gov/oag/newsreleases15/ pr20150526b.html, 2015.

\bibitem{digital_ad_vs_tv}
P.~Kafka and R.~Molla.
\newblock {Recode - 2017 was the year digital ad spending finally beat TV}.
\newblock https://www.recode.net/2017/12/4/16733460/
  2017-digital-ad-spend-advertising-beat-tv, 2017.

\bibitem{nominer}
R.~Keramidas.
\newblock No coin.
\newblock https://github.com/keraf/NoCoin.

\bibitem{incomeExample}
H.~Lau.
\newblock Browser-based cryptocurrency mining makes unexpected return from the
  dead.
\newblock
  https://www.symantec.com/blogs/threat-intelligence/browser-mining-cryptocurrency.

\bibitem{miningSites}
J.~Leyden.
\newblock More and more websites are mining crypto-coins in your browser to pay
  their bills, line pockets.
\newblock https://www.theregister.co.uk/2017/10/13/ crypto\_mining/.

\bibitem{ronaldo}
J.~Leyden.
\newblock Real mad-quid: Murky cryptojacking menace that smacked ronaldo site
  grows.
\newblock http://www.theregister.co.uk/2017/10/10/ cryptojacking/.

\bibitem{commissioner}
N.~Lomas.
\newblock Cryptojacking attack hits 4,000 websites, including uk’s data
  watchdog.
\newblock https://techcrunch.com/2018/02/12/ico-snafu/, 2018.

\bibitem{optimalTemp}
J.~Martin.
\newblock What's the best cpu temperature?
\newblock https://www.techadvisor.co.uk/how-to/desktop-pc/cpu-temp-3498564/,
  2018.

\bibitem{showtime}
K.~McCarthy.
\newblock Cbs's showtime caught mining crypto-coins in viewers' web browsers.
\newblock http://www.theregister.co.uk/2017/09/25/
  showtime\_hit\_with\_coinmining\_script/.

\bibitem{websocket}
{MDN web docs}.
\newblock Websockets.
\newblock https://developer.mozilla.org/en-US/docs/Web/API/WebSockets\_API.

\bibitem{opera-nocoin}
K.~Mielczarczyk.
\newblock Opera 50 beta rc with cryptocurrency mining protection.
\newblock
  https://blogs.opera.com/desktop/2017/12/opera-50-beta-rc-cryptocurrency-mining-protection/,
  2017.

\bibitem{adlabels}
C.~Morran.
\newblock Why do websites refuse to label sponsored content as
  “advertising”?
\newblock
  https://consumerist.com/2015/06/11/why-do-websites-refuse-to-label-sponsored-content-as-advertising/.

\bibitem{forbes}
B.~Morrissey.
\newblock Forbes starts blocking ad-block users.
\newblock https://digiday.com/media/forbes-ad-blocking/.

\bibitem{web-workers}
{Mozilla Developers}.
\newblock Web workers api.
\newblock https://developer.mozilla.org/en-US/docs/Web/API/Web\_Workers\_API.

\bibitem{mughees2017detecting}
M.~H. Mughees, Z.~Qian, and Z.~Shafiq.
\newblock Detecting anti ad-blockers in the wild.
\newblock {\em Proceedings on Privacy Enhancing Technologies},
  2017(3):130--146, 2017.

\bibitem{revenueDrop}
M.~H. Mughees, Z.~Qian, Z.~Shafiq, K.~Dash, and P.~Hui.
\newblock A first look at ad-block detection: {A} new arms race on the web.
\newblock {\em CoRR}, abs/1605.05841, 2016.

\bibitem{armsRace}
A.~Muller.
\newblock Ad-mageddon! ad blocking, its impact, and what comes next.
\newblock
  https://marketingland.com/ad-mageddon-perspectives-ad-blocking-impacts-comes-next-227090.

\bibitem{bad-packets-report}
T.~Mursch.
\newblock Cryptojacking: 2017 year-end review.
\newblock https://badpackets.net/cryptojacking-2017-year-end-review/, 2017.

\bibitem{bitcoinPlus}
D.~Nadolny.
\newblock Bitcoin plus miner.
\newblock https://wordpress.org/plugins/bitcoin-plus-miner/.

\bibitem{userTime}
J.~NIELSEN.
\newblock How long do users stay on web pages?
\newblock
  https://www.nngroup.com/articles/how-long-do-users-stay-on-web-pages/.

\bibitem{Nikiforakis:2013:CME:2497621.2498133}
N.~Nikiforakis, A.~Kapravelos, W.~Joosen, C.~Kruegel, F.~Piessens, and
  G.~Vigna.
\newblock Cookieless monster: Exploring the ecosystem of web-based device
  fingerprinting.
\newblock In {\em Proceedings of the IEEE Symposium on SP '13}.

\bibitem{nithyanand2016adblocking}
R.~Nithyanand, S.~Khattak, M.~Javed, N.~Vallina-Rodriguez, M.~Falahrastegar,
  J.~E. Powles, E.~D. Cristofaro, H.~Haddadi, and S.~J. Murdoch.
\newblock Adblocking and counter blocking: A slice of the arms race.
\newblock In {\em 6th {USENIX} Workshop on Free and Open Communications on the
  Internet ({FOCI} 16)}, Austin, TX, 2016. {USENIX} Association.

\bibitem{nomining}
{No Mining}.
\newblock Secure your browser with no mining.
\newblock http://www.nomining.com/.

\bibitem{gdpr}
{Official Journal of the European Union}.
\newblock Directive 95/46/ec (general data protection regulation).
\newblock
  http://eur-lex.europa.eu/legal-content/EN/TXT/PDF/?uri=CELEX:32016R0679.

\bibitem{browselowd}
P.~Paganini.
\newblock Thousands of websites worldwide hijacked by cryptocurrency mining
  code due browsealoud plugin hack.
\newblock https://securityaffairs.co/wordpress/68966/
  hacking/browsealoud-plugin-hack.html, 2018.

\bibitem{appsVsWeb}
E.~P. Papadopoulos, M.~Diamantaris, P.~Papadopoulos, T.~Petsas, S.~Ioannidis,
  and E.~P. Markatos.
\newblock The long-standing privacy debate: Mobile websites vs mobile apps.
\newblock In {\em Proceedings of the 26th International Conference on World
  Wide Web}, WWW'17, 2017.

\bibitem{www18adcost}
P.~Papadopoulos, N.~Kourtellis, and E.~P. Markatos.
\newblock The cost of digital advertisement: Comparing user and advertiser
  views.
\newblock In {\em Proceedings of the 27th International Conference on World
  Wide Web}, WWW'18, 2018.

\bibitem{imcRTB}
P.~Papadopoulos, N.~Kourtellis, P.~R. Rodriguez, and N.~Laoutaris.
\newblock If you are not paying for it, you are the product: How much do
  advertisers pay to reach you?
\newblock In {\em Proceedings of the 2017 Internet Measurement Conference}, IMC
  '17, 2017.

\bibitem{kaspersky}
A.~Perekalin.
\newblock How kaspersky lab products protect against miners.
\newblock https://www.kaspersky.com/blog/web-miners-protection/20556/, 2018.

\bibitem{phidget1}
{Phidgets Inc.}
\newblock Phidget21 c api documentation.
\newblock https://www.phidgets.com/documentation/web/ cdoc/index.html.

\bibitem{phidget2}
{Phidgets Inc.}
\newblock What is a phidget?
\newblock https://www.phidgets.com/docs21/ What\_is\_a\_Phidget.

\bibitem{optimalTemp2}
M.~Pinola.
\newblock How to test your computer's cpu temperature.
\newblock https://www.lifewire.com/how-can-i-test-laptop-temperature-2377618,
  2018.

\bibitem{publicWWW}
{PublicWWW}.
\newblock Source code search engine.
\newblock https://publicwww.com/.

\bibitem{razaghpanah2018apps}
A.~Razaghpanah, R.~Nithyanand, N.~Vallina-Rodriguez, S.~Sundaresan, M.~Allman,
  C.~Kreibich, and P.~Gill.
\newblock Apps, trackers, privacy, and regulators: A global study of the mobile
  tracking ecosystem.
\newblock 2018.

\bibitem{psrecord}
T.~Robitaille.
\newblock psrecord: Record the cpu and memory activity of a process.
\newblock https://github.com/astrofrog/psrecord.

\bibitem{lmsensors}
G.~Roeck.
\newblock Overview of the lm-sensors package.
\newblock https://github.com/groeck/lm-sensors.

\bibitem{monero}
A.~Rosic.
\newblock What is monero? the ultimate beginners guide.
\newblock https://blockgeeks.com/guides/monero/.

\bibitem{webminingwaste}
K.~Sedgwick.
\newblock {Mining Crypto In a Browser Is a Complete Waste of Time}.
\newblock
  https://news.bitcoin.com/mining-crypto-in-a-browser-is-a-complete-waste-of-time/,
  2018.

\bibitem{evading-blacklists}
J.~Segura.
\newblock Malicious cryptomining and the blacklist conundrum.
\newblock
  https://blog.malwarebytes.com/threat-analysis/2018/03/malicious-cryptomining-and-the-blacklist-conundrum/,
  2018.

\bibitem{obfuscation}
D.~Sinegubko.
\newblock Malicious website cryptominers from github.
\newblock
  https://blog.sucuri.net/2018/01/malicious-cryptominers-from-github-part-2.html,
  2018.

\bibitem{financialTimes}
N.~Smith.
\newblock How publishers are turning up the heat in the ad-blocking war.
\newblock
  https://www.theguardian.com/media-network/2016/sep/02/publishers-ad-block-users-hide-content.

\bibitem{pow}
A.~Tar.
\newblock Proof-of-work, explained.
\newblock https://cointelegraph.com/explained/proof-of-work-explained, 2018.

\bibitem{wired}
{The Editors of Wired}.
\newblock How wired is going to handle ad blocking.
\newblock https://www.wired.com/how-wired-is-going-to-handle-ad-blocking/.

\bibitem{malvertising}
{The European Union Agency for Network and Information Security (ENISA)}.
\newblock Malvertising.
\newblock https://www.enisa.europa.eu/publications/info-notes/malvertising,
  2016.

\bibitem{ps}
{The Linux Information Project}.
\newblock The ps command.
\newblock http://www.linfo.org/ps.html.

\bibitem{programmaticAd}
K.~Tse.
\newblock Understanding programmatic advertising: A brief look at its history.
\newblock
  https://medium.com/wired-mesh/understanding-programmatic-advertising-a-brief-look-at-its-history-411dd5842304.

\bibitem{popunder}
L.~Tung.
\newblock Windows: This sneaky cryptominer hides behind taskbar even after you
  exit browser.
\newblock
  https://www.zdnet.com/article/windows-this-sneaky-cryptominer-hides-behind-taskbar-even-after-you-exit-browser/,
  2017.

\bibitem{breaking}
N.~Vallina-Rodriguez, J.~Shah, A.~Finamore, Y.~Grunenberger, K.~Papagiannaki,
  H.~Haddadi, and J.~Crowcroft.
\newblock Breaking for commercials: Characterizing mobile advertising.
\newblock In {\em Proceedings of the 2012 ACM Conference on Internet
  Measurement Conference}, IMC '12, 2012.

\bibitem{vallina2016tracking}
N.~Vallina-Rodriguez, S.~Sundaresan, A.~Razaghpanah, R.~Nithyanand, M.~Allman,
  C.~Kreibich, and P.~Gill.
\newblock Tracking the trackers: Towards understanding the mobile advertising
  and tracking ecosystem.
\newblock {\em arXiv preprint arXiv:1609.07190}, 2016.

\bibitem{blocking-extensions}
A.~Verma.
\newblock 6 easy ways to block cryptocurrency mining in your web browser.
\newblock https://fossbytes.com/block-cryptocurrency-mining-in-browser/, 2018.

\bibitem{opera-mobile}
A.~Viquez.
\newblock Opera introduces bitcoin mining protection in all mobile browsers –
  here’s how we did it.
\newblock
  https://blogs.opera.com/mobile/2018/01/opera-introduces-bitcoin-mining-protection-mobile-browsers/,
  2018.

\bibitem{chrome-extensions}
J.~Wagner.
\newblock Protecting users from extension cryptojacking.
\newblock
  https://blog.chromium.org/2018/04/protecting-users-from-extension-cryptojacking.html,
  2018.

\bibitem{webassembly}
M.~web docs.
\newblock Webassembly.
\newblock https://developer.mozilla.org/en-US/docs/WebAssembly.

\bibitem{dataplan3}
{WhistleOut Inc.}
\newblock Compare the best cell phone plans.
\newblock https://www.whistleout.com/CellPhones, 2018.

\bibitem{adblockplus}
B.~Williams.
\newblock Adblock plus and (a little) more.
\newblock
  https://adblockplus.org/blog/100-million-users-100-million-thank-yous.

\bibitem{sophos}
J.~Wyke.
\newblock The zeroaccess botnet: Mining and fraud for massive financial gain.
\newblock Sophos Technical Paper, 2012.

\bibitem{Zarras:2014:DAM:2663716.2663719}
A.~Zarras, A.~Kapravelos, G.~Stringhini, T.~Holz, C.~Kruegel, and G.~Vigna.
\newblock The dark alleys of madison avenue: Understanding malicious
  advertisements.
\newblock In {\em Proceedings of the 2014 Conference on Internet Measurement
  Conference}, IMC '14, pages 373--380, New York, NY, USA, 2014. ACM.

\bibitem{coinblockerlists}
{ZeroDot1}.
\newblock Coinblockerlists.
\newblock https://github.com/ZeroDot1/CoinBlockerLists.

\end{thebibliography}
